\documentclass[prb,floatfix]{revtex4}
\usepackage{graphics}
\usepackage{subfigure}
\usepackage{epsfig}

\newcommand{\ket}[1]{|#1\rangle}
\newcommand{\bra}[1]{\langle #1|}

\begin{document}

\title{Implementing a Quantum Algorithm with Exchange-Coupled Quantum
Dots: a Feasibility study.}
\author{E. Simon Myrgren and K. Birgitta Whaley}
\affiliation{Department of Chemistry and Kenneth S. Pitzer Center for
Theoretical Chemistry, \\ University of California, Berkeley, CA
94720, USA}
\date{\today}

\begin{abstract}
We present Monte Carlo wavefunction simulations for quantum
computations employing an  
exchange-coupled array of quantum dots.  Employing a combination of
experimentally and theoretically available parameters, we find that
gate fidelities greater than 98 \% may be obtained with
current experimental and technological capabilities.  Application to an encoded 3 qubit 
(nine physical qubits) Deutsch-Josza computation indicates that the algorithmic
fidelity is more a question of the total time to implement the gates
than of the physical complexity of those gates. 
\end{abstract}

\maketitle

\section{Introduction}
\label{sec:intro}
Quantum information processing has been characterized by a rapid
pace of theoretical progress and slower development of experimental
techniques. At the current time, the
differential is clearly  
visible: while most experimental realizations to date are limited to a
handful of qubits or less, much theoretical effort is devoted to
elucidating the potential of quantum computers with thousands or even
millions of qubits. As experiments progress there is a need to
evaluate the many suggested experimental implementations, to determine
if they are feasible as 
proposed. The evaluations need not be elaborate, just the calculation
of some figures of merit that indicate the technology's computational
merit. As an example of such a calculation, we present here
simulation results and fidelities for the CNOT gate and for the three
qubit Deutsch-Josza algorithm implemented in a model system of 
a spin-coupled quantum dot array with exchange-only quantum
computation.$^($\cite{divincenzo00,kempe01,bacon00}$^)$ 

The possibility of employing coupled quantum wells for
quantum information processing was first proposed by Landauer, and
Barenco {\em et al.} in the mid-1990's.$^($\cite{landauer96,barenco95}$^)$ Numerous
researchers have since elaborated upon these ideas. 
We investigate here the quantum dot
implementation proposed by Loss and DiVincenzo.$^($\cite{loss98}$^)$ 
As suggested by these authors, we consider a linear quantum dot
array where each dot arises 
as a localized region within a two dimensional electron gas, with the localization
imposed by electrical gating. Each qubit is
realized as the spin of an unpaired electron on the quantum
dot. Following Ref.~\onlinecite{divincenzo98}, we shall assume the spin-orbit 
interaction is negligible,
and the effect of surrounding nuclear spins will be incorporated into the
error terms. Thus, the spin of each electron constitutes a well-defined two
dimensional Hilbert space. Employing spin rather than
orbital degrees of freedom greatly reduces the effects of decoherence,
since the 
spin states couple much less strongly to the environment than the
charge states.$^($\cite{burkard99}$^)$ 
Whereas the charge degrees of freedom are
characterized by a decoherence rate on the order of
nanoseconds$^($\cite{kikkawa97}$^)$, the 
spin degrees are relatively resistant to errors, and dephasing rates
in the microsecond range can be expected.$^($\cite{fujisawa01}$^)$
It should 
be noted that these figures stem from experiments done under
non-decoherence suppressing conditions, and one might expect that
with more elaborate experimental set-ups, {\em e.g.} taking advantage
of spin polarization and spin echo techniques, the effects of
inhomogeneous broadening as well as the hyperfine coupling between the
electron and surrounding nuclear spins can be reduced. To date,
there has been no actual experimental realization of the spin-coupled 
quantum dot array. However, demonstrated
experimental ability to control the coupling between the
dots$^($\cite{livermore96,tarucha96}$^)$ and
observations of coherent and long-lived spin oscillations$^($\cite{gupta99}$^)$
indicates that a quantum computer as envisioned by Loss and DiVincenzo
could be realizable.

Assuming that tunable gates, well-defined arrays, and reasonable
control of decoherence processes have been achieved, the issue of
measurement remains. Less work has been done on
the realization of quantum measurement in these systems, although
efforts towards experimental realization of single
spin measurements in the solid state have been
made.$^($\cite{wiseman01,engel01,divincenzo99b}$^)$ Since at the outset of
these simulations no experimentally demonstrated 
measurement scheme existed, we have made the simplest assumption of
noiseless projective measurements.

In this work we employ the isotropic exchange interaction for coupling
quantum spins with the exchange-only quantum computation scheme of
Refs.~\onlinecite{divincenzo00},\onlinecite{kempe01}, and
\onlinecite{bacon00}. Use of an isotropic 
exchange interaction amounts to an idealization
of the system as it generally exists in an experimental setting.
For real quantum dots, some exchange anisotropy may arise from the
effects of the finite
spin-orbit coupling.$^($\cite{loss98}$^)$ 
This anisotropy may be dealt with by modified
encoding$^($\cite{kempe01,vala02}$^)$ or by pulse-shaping
techniques.$^($\cite{bonesteel01}$^)$ Both of these approaches can in
principle be applied to the simulation analysis presented here.

\section{Universality through exchange}

\label{sec:exchange}
A criteria for any quantum processing device is that two qubits can be
coupled in a non-trivial way. In the quantum dot arrays under
consideration here, this coupling arises from the nearest neighbour spin-spin
interactions described by the Heisenberg exchange
Hamiltonian:$^($\cite{loss98}$^)$  
\begin{equation}
H_{ij} = J(t) {\bf S}_i \cdot {\bf S}_j,  \label{Heisen}
\end{equation}
where ${\bf S}_i= 1/2 \left( \sigma_{ix}, \sigma_{iy},
\sigma_{iz} \right)$ are
the spin operators at site $i$, with $\sigma_{i\alpha}$ the Pauli matrices, and
$J(t)$ is the exchange coupling strength. Loss and DiVincenzo have
shown that when the exchange interaction is 
pulsable (i.e., it can be switched between finite ``on'' and low or
zero ``off'' values) this exchange interaction 
allows the implementation of 
$\sqrt{SWAP}$, which, in conjunction with local unitaries, is
equivalent to XOR (also known as CNOT).$^($\cite{loss98}$^)$ 
The switching is controlled by the integrated coupling strength
$\int J(t) dt = J_0 t_s$ over the pulse duration $t_s$, with the 
$SWAP$ gate being achieved for $J_0 t_s = \pi(mod2\pi)$ and
$\sqrt{SWAP}$ for one half of this.
To facilitate calculations with this spin Hamiltonian it is convenient
to rescale it by addition of a unit operator, to arrive at the
exchange operator $E_{ij}$
\begin{eqnarray}
E_{ij} & = &  2 ( {\bf{S}}_i \cdot {\bf{S}}_j + \frac{1}{4}{\bf{I}}_4 )
\nonumber \\ 
       & = &  \frac{1}{2}\left( \sigma_i \otimes \sigma_j + I_i
\otimes I_j \right), 
\label{eqn:heisenbergtoexchange}
\end{eqnarray}
where  $\bf{I}_4$ is the 4-by-4 identity matrix for
the two-qubit Hilbert space.
This rescaled operator acts to exchange the states of physical qubits
$i$ and $j$. In the rest of this paper all exchange gate
times pertaining to single or two qubit gates will be given based on
this exchange operator, unless otherwise indicated. As such they will be
expressed in units of $2\hbar/J_0$.   

\begin{figure}[ht] 
\centering
\includegraphics[width=4in, height=2in]{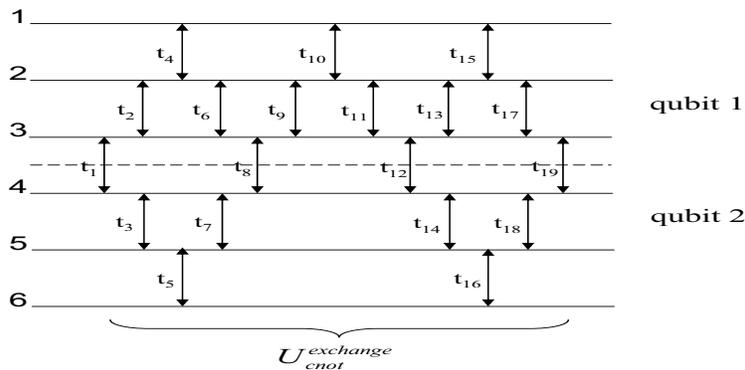}
\caption{Circuit diagram for implementing a CNOT gate between logical
qubits 1 and 2. Logical qubit 1 is encoded in the top three physical
qubits, 1-3, and  
logical qubit 2 in the bottom three physical qubits, 4-6.  Each
vertical line connecting  
two physical qubits represents an 
exchange gate with strength $J_0$ switched on for time $t_s$. As a result of the symmetry of the gate sequence, only
7 independent  
variables are required to define the gate times of the entire 19-gate
sequence.$^($\cite{divincenzo00}$^)$} 
\label{fig:fig1}
\end{figure}

Unfortunately, in
the solid state one-qubit gates are often technically more demanding
than two-qubit 
gates.$^($\cite{divincenzo00}$^)$ Thus even the simple one-qubit gates required to
obtain the XOR from $\sqrt{SWAP}$ require a high degree of experimental
and technical sophistication. 
Kempe {\em et al.} showed that the Heisenberg interaction can be made 
universal by itself, with the use of a suitable encoding (``encoded 
universality'').$^($\cite{kempe01,divincenzo00,vala02}$^)$ For a linear
quantum dot array 
as that considered here, one possible encoding is to represent
each logical qubit using three physical qubits:
\begin{eqnarray}
\ket{0}_L & =  & \frac{1}{\sqrt{2}} ( \ket{011} - \ket{101} ),
\nonumber \\ 
\ket{1}_L & = & \sqrt{\frac{2}{3}} \ket{110} - \frac{1}{\sqrt{6}} \
( \ket{011} + \ket{101} ).
\label{eqn:encoding}
\end{eqnarray}    
Note that total spin along the z-axis, $S_z^{tot}=\sum_i S_{i,z}$, is
conserved and equal to $\frac{1}{2}$, and that the states, in
addition, are eigenstates of $S^2$. 
The computational basis is $\ket{00}_L, \, \ket{01}_L, \, \ket{10}_L$,
and $\ket{11}_L$.  
With this encoding, an $SU(4)$ operation that is locally equivalent to 
CNOT between logical qubits, has been shown to be feasible with 19 
exchanges$^($\cite{divincenzo00}$^)$ between adjacent pairs of physical
qubits.  This operation is illustrated in Fig.~\ref{fig:fig1}.

It should be noted that 19 is an upper limit on
the number of serial exchanges required, obtained through numerical
optimization of the Makhlin invariants$^($\cite{makhlin02}$^)$, and it is possible
that a smaller number might suffice.
The number of exchange operations can also 
be reduced if one considers a more complex quantum dot architecture
that supports
non-nearest neighbour interactions, rather than the linear array
studied here.  In
particular, if in addition to exchanges $H_{ii+1}$, any $H_{ij}$
connection is accessible, then full parallelism is possible. With parallel
exchanges of 
this type a one-qubit rotation can be shown to require only  
3 exchanges, and a CNOT gate only 8 exchanges.$^($\cite{divincenzo00}$^)$ 

\begin{figure}[ht]
\centering
\includegraphics[width=4in, height=1.25in]{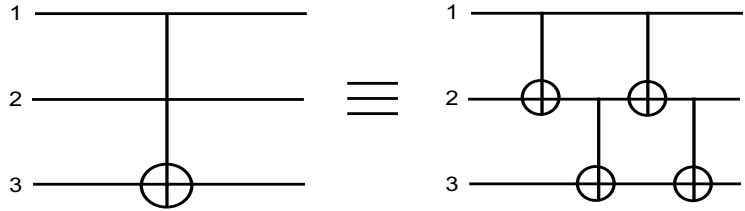}
\caption{Equivalence of circuit employing only CNOT gates
between adjacent logical qubits, to a circuit representing a CNOT
between qubits 1 and 3.}
\label{fig:fig2}
\end{figure}

Universal quantum information processing will require that we
can implement the CNOT gate between any two logical 
qubits, and not just between adjacent qubits. This need is apparent even
for such simple
algorithms as the three qubit Deutsch-Josza algorithm that tests whether
an arbitrary function 
is balanced or constant.$^($\cite{nielsen00}$^)$ We shall see below that one
of the black box gate sequences describing a balanced function
for the three qubit Deutsch-Josza algorithm requires a CNOT gate
between qubits 1 and 3. We shall refer to this as CNOT(1,3).  
Analysis of truth tables for combinations of
CNOTs between neighbouring dots readily shows that
a CNOT(1,3) gate is equivalent to a pairwise sequence of four CNOT
gates between adjacent 
dots, as illustrated in Fig.~\ref{fig:fig2} 

It would be perfectly valid to employ this
pairwise adjacent sequence to represent CNOT(1,3). However, since each
CNOT requires  
19 exchange operations, this would entail implementing a total of
$4*19=76$ exchange 
gates. Here, recognition of the relation between exchange and SWAP
operations allows for a more  
efficient solution. We note that
any exchange gate when applied for a duration equivalent to a
$\pi$ pulse, yields the SWAP operation, and that 
exchange gates may be inter converted by action of the appropriate SWAP
operations, {\em e.g.}, 
\begin{equation}
\exp{(it H_{13} / \hbar )}=\exp{( i \pi H_{23} / \hbar )} \exp{( i t
    H_{12} / \hbar )} \exp{(i \pi H_{23} / \hbar )}.  
\end{equation}
Combining an exchange gate between logical qubits $i$ and
$i+1$ with exchange-generated SWAP operations immediately before and
after, thereby makes the exchange operational between 
any two qubits $i$ and $j$. This is summarized by the following: 
\begin{equation}
\exp{(itH_{ij}/ \hbar)} = \prod_{k=i+1}^{j-1} \exp{ \left( i \pi
        H_{kk+1} / \hbar  \right) } 
        \exp{ \left( itH_{ii+1} / \hbar \right) } \prod_{k=j-1}^{i+1}
        \exp{ \left( i\pi H_{kk+1} / \hbar \right)} .  
\end{equation}           
Making use of this relation, we find that a
CNOT gate between qubits 1 and 3 can now be performed with only 55 exchanges,  
an overall saving of 21 operations compared with the pairwise adjacent
scenario depicted in Fig~\ref{fig:fig2}.   

\subsection{The exact CNOT}
\label{subsec:exactcnot}
The Makhlin invariants$^($\cite{makhlin02}$^)$ guarantee a sequence of
exchange gates and times that provide a two-qubit unitary $U_{cnot}^{exchange}$
which is locally equivalent to the 
CNOT. This is not necessarily equal to CNOT in the computational
basis. To use our 
exchange-only CNOT in conjunction with other gates, we 
therefore need to find a set of local unitaries that transform this
locally equivalent gate to the CNOT in the computational basis.
Mathematically, we can represent the transformation as:
\begin{equation}
CNOT \, = \, \left( U_1 \otimes U_2 \right) \, U_{cnot}^{exchange} \\ 
\, \left( V_1 \otimes V_2 \right).
\label{eq:CNOT_exchange}
\end{equation}
Here $U_1$, $U_2$, $V_1$, and $V_2$ designate local basis
transformations each consisting of at most 4 exchange
gates$^($\cite{divincenzo00}$^)$ that act on the 
first or the second logical qubit. For $U_{cnot}^{exchange}$ we employ
here the optimized sequence of 19 exchange gates from
Ref.~\onlinecite{divincenzo00}. It is then possible, using a procedure
introduced by Makhlin$^($\cite{makhlin02}$^)$, to find the local
unitaries ($U_1$, $U_2$, $V_1$, and $V_2$) by 
i) recasting $U_{cnot}^{exchange}$ in the Bell basis,
$M_B=Q^{\dagger}U_{cnot}^{exchange}Q$, ii) evaluating the spectrum
of $m=M_B^TM_B$, and iii) then relating this matrix of eigenvalues to the
corresponding matrix for the CNOT gate in the computational basis. 
Here $Q$ is the matrix that
transforms $U_{cnot}^{exchange}$ from the computational basis to the Bell basis, and 
$M_B^T$ is the matrix transpose of $M_B$. Having obtained the local
unitaries $U_i$ and $V_i$, $i=1,2$, it remains to decompose these into an actual sequence of
exchange gates in order to perform full exchange-only computation. We
describe here two ways to accomplish this decomposition. The first is
a general procedure based on numerical optimization. The second is an
analytic procedure that is specific to the present case, since it
relies on the ability to find analytic solutions to systems of
trigonometric equations that may be less tractable for other situations.

For both approaches we give the times $t_s$ for the individual exchange
gates in units of $2 \hbar / J_0$. We cite all time values as positive
numbers here, and implicitly assume that the pulse-integrated exchange 
coupling value $J_0$ is constant. (In Ref.~\onlinecite{burkard99},
Burkard {\em et al.} consider the possibility of tuning the Heisenberg
interaction around the point of zero coupling, thus allowing $J_0$ to assume
both positive and negative values.) 
Note that all gate times are defined modulo $\pi$,
{\em i.e.} an exchange gate implemented for $t_s$ is identical (up to a 
global phase of -1) to one
implemented for $t_s+\pi$ (see discussion above and
Refs. \onlinecite{loss98} and \onlinecite{divincenzo00}).

For the numerical optimization procedure, we first express the local basis 
transformations, $U_1$, $U_2$, $V_1$, and
$V_2$,  as sequences of 4 exchange gates. We then
use a version of the Nelder-Mead simplex
algorithm$^($\cite{nelder65}$^)$ to find those sequences that minimize both
the matrix distance from the resulting CNOT gate Eq.~(\ref{eq:CNOT_exchange}) 
to the
true CNOT in the computational basis, and the extent of leakage out of the
encoded subspace.
The Nelder-Mead algorithm is an example of a direct search method, {\em i.e.} 
it uses only function evaluations and does not rely upon any derivative
information about the cost function. Each iteration begins with 
a geometric figure, a {\em simplex}, created from $m+1$
coordinates in parameter space, where $m$ is the number of variables
of the cost function to be optimized. From this first simplex, new points
are generated and the cost function is evaluated at these new
coordinates. A new simplex, possessing better descent characteristics than the
previous, is then generated from the cost function evaluations and the new
test points.$^($\cite{lagarias95}$^)$
The Nelder-Mead method is known to work well in low-dimensional
instances such as those studied here. The local character of the method
is nevertheless a concern. To avoid getting trapped in local minima, the
parameter space must be densely sampled. We accomplish this here by
{\em shooting} initial coordinates into parameter space, followed by
optimization from the coordinate that gave the smallest initial value
of the cost function. 

\begin{figure}[ht]
\centering
\includegraphics[width=3.5in, height=2.5in]{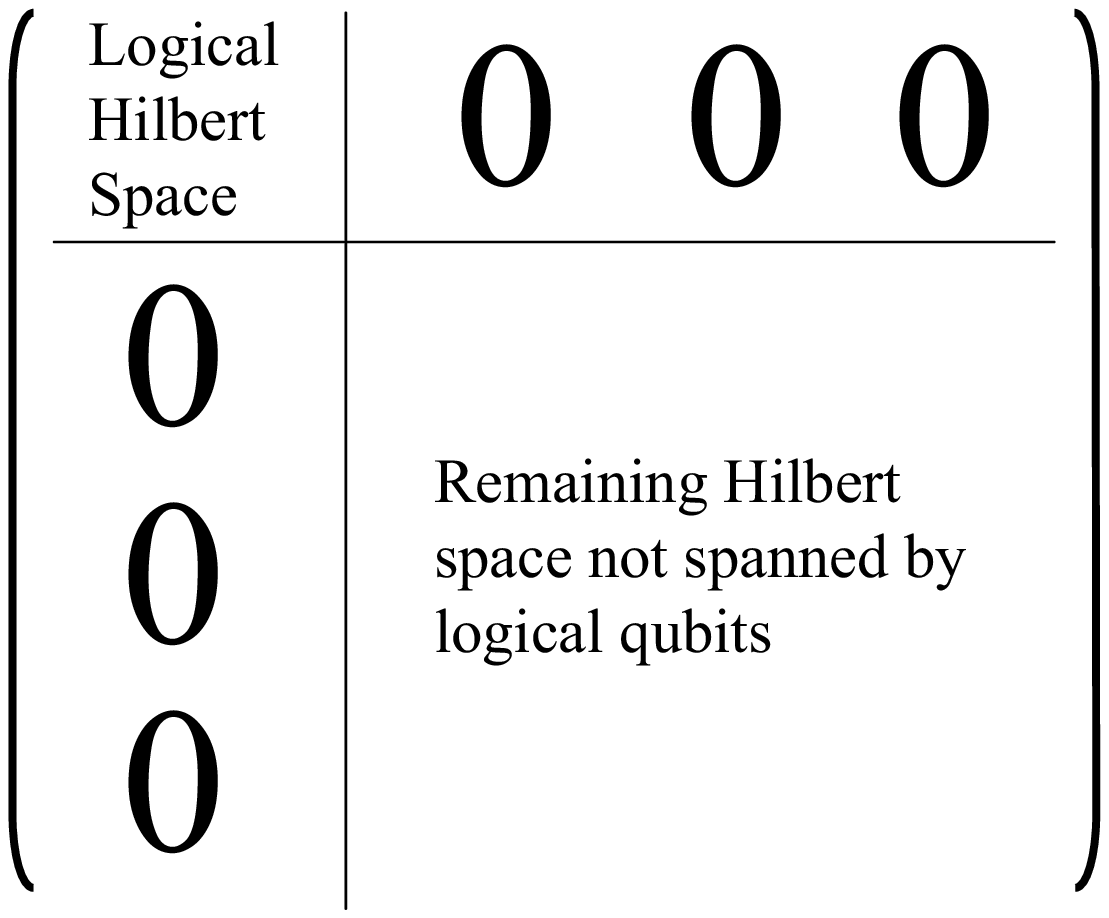}
\caption{To ensure that our encoded unitary does not leak, we require
that the encoded unitary, $\left( U_1 \otimes U_2 \right) \,
U_{cnot}^{exchange} \, \left( V_1 \otimes V_2 \right)$, can be
represented by the above matrix decomposition. The logical
Hilbert space has dimension $2^n \times 2^n$ for n logical qubits. The
complete Hilbert space has dimension $2^N \times 2^N$, corresponding to N
physical qubits.} 
\label{fig:fig3}
\end{figure}

As noted above, our cost function contains two
components. First, it includes an element by 
element matrix equivalence criteria, {\em i.e.} the matrix distance
between the target gate (the true CNOT) and the candidate exchange-only
representation of this, 
Eq.~(\ref{eq:CNOT_exchange}), which is constructed from the 19-exchange 
sequence of Ref.~\onlinecite{divincenzo00} together with two 8-exchange
sequences representing the local unitaries.  Second, it contains a component 
that provides a penalty for leakage out of the encoded subspace,
namely the sum of the absolute value of all matrix elements connecting
states in the encoded logical subspace to states outside this
subspace, illustrated in Fig.~\ref{fig:fig3}.

These two requirements of minimal matrix distance from the target CNOT 
and non-leakage from the encoded subspace lead to the following expression for
the cost function: 
\begin{equation}
C= \sum_{i,j}^{2^n} | U_{cnot}^{exchange}(i,j) - CNOT(i,j) | + \ 
        \sum_{2^n < i < 2^N \atop j \le 2^n} \
        | U_{cnot}^{exchange}(i,j) |,
\label{eqn:costfunction}
\end{equation} 
where $n$ is the number of logical qubits and $N$ the number of physical
qubits. The first summation
represents the matrix distance between Eq.~(\ref{eq:CNOT_exchange}) and the
target CNOT, and the second summation, which goes over
the lower left-hand block of the matrix in Fig.~\ref{fig:fig3}, represents the
extent of leakage.  The second summation thus consists of terms connecting
basis states $j \leq 2^n$ inside the encoded logical subspace to all
basis states $2^n < i \leq 2^N$ outside the encoded logical subspace.

\begin{figure}[ht]
\centering
\includegraphics[width=3in, height=2in]{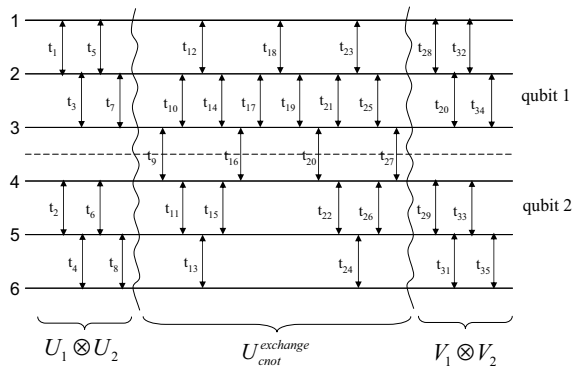}
\caption{Diagrammatic representation of one set of exchange gates
that transform the 19-exchange gate CNOT ($U_{cnot}^{exchange}$) into the exact
CNOT. These 8+8 gate times ($t_1$-$t_8$ and $t_{28}$-$t_{35}$) were
arrived at numerically using 
the Nelder-Mead  simplex method$^($\cite{nelder65}$^)$ as described in the
text.  The
corresponding 35 gate times given in Table~\ref{tab:tab1}
will generate the exact CNOT to within a cost function value
$C=10^{-5}$, corresponding to a maximum matrix element distance of
$5 \times 10^{-6}$. } 
\label{fig:fig4}
\end{figure}

Since we employ the 19-exchange sequence for $U_{cnot}^{exchange}$ from
Ref.~\onlinecite{divincenzo00}, the numerical optimization is restricted 
here to the two sets of 8-exchange sequences representing $U_1\otimes U_2$ and
$V_1 \otimes V_2$, respectively.  We implement this by
constructing a sequence of 35 gates (i.e., 4+4+19+4+4, see Fig.~\ref{fig:fig4})
with gates 9-27 taken from Ref.~\onlinecite{divincenzo00}, and then 
optimizing the
cost function $C$, Eq.~(\ref{eqn:costfunction}), for the matrix resulting from
the entire 35-exchange sequence only over the 16-dimensional parameter space of the 
two 8-exchange sequences.   We note that since the 
available nearest neighbor exchanges $E_{12}$ and
$E_{23}$ correspond simply to rotations of the logical states around the 
z-axis and about an axis
oriented along $\sqrt{3}/2 \sigma_x + 1/2 \sigma_z$,
respectively$^($\cite{kempe01,divincenzo00}$^)$, the exchange-based local gates 
will therefore not take states outside
the logical subspace and will hence not add to the leakage term.  The leakage
parameter is therefore determined solely by the accuracy of the
underlying gate sequence for $U_{cnot}^{exchange}$.  

We have found that for such a 35-gate sequence, the overall
cost function C can readily be reduced to less than $10^{-4}$ by this 
numerical optimization.  More specifically, we find that the matrix
distance (first term of Eq.~\ref{eqn:costfunction}) between this 35
gate long sequence and the exact CNOT is $1\times 10^{-5}$ and that the 
leakage term (second term of Eq.~\ref{eqn:costfunction}) is $5 \times
10^{-9}$ (equal to the value for $U_{cnot}^{exchange}$, as noted above). 
This level of optimization corresponds to a maximum element 
matrix distance from CNOT ({\it i.e.}, maximum matrix element inaccuracy) of 
$5\times 10^{-6}$.  
The optimal exchange-only sequences for the local transformations are 
summarized in Fig.~\ref{fig:fig4} and in
Table~\ref{tab:tab1}, respectively. 

\begin{figure}[ht]
\centering
\subfigure[
A single qubit gate
($\frac{\sqrt{3}}{2}-\frac{i}{2}\sigma_y$) acting on the second
logical qubit diagonalizes the 19-gate exchange
sequence. 
The resulting diagonal 4-by-4 matrix is
then converted into the C-PHASE by $\sigma_z$-rotations acting on both
the first and the second qubit, with angles $\phi=0.612497$ and
$\theta=-0.547580$, respectively. These values are determined 
from the analytic solutions 
to a linear equation system with 3 unknowns: $\phi$, $\theta$ and a
global phase. See Appendix~\ref{app:quaternions} for details as to how these
parameters were obtained.]{
  \label{fig:fig5a}
  \includegraphics[width=4in,height=1in]{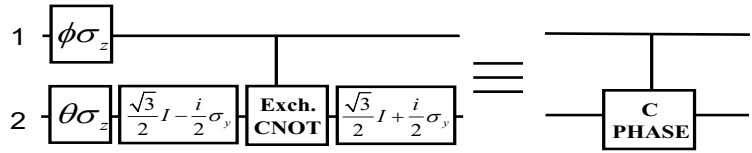}}
\vspace{.2in}
\subfigure[The C-PHASE gate can be transformed into the CNOT
gate by acting with Hadamard gates on the second qubit before and after the
C-PHASE gate.]{
  \label{fig:fig5b}
  \includegraphics[width=4in,height=1in]{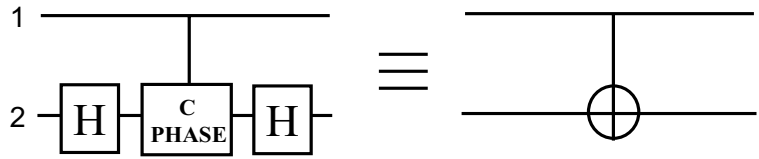}}
\caption{Representation of analytic sequence of local transformations
that transform the 19-exchange sequence $U_{cnot}^{exchange}$ from 
Ref.~\onlinecite{divincenzo00} into the true CNOT in the computational basis. 
The exchange gates and times
corresponding to the elementary local transformations are then 
obtained using the quaternion representation of the desired $SU(2)$
unitaries (see Appendix~\ref{app:quaternions} for details).}
\label{fig:fig5} 
\end{figure}

\begin{figure}[ht]
\centering
\includegraphics[width=3in, height=2in]{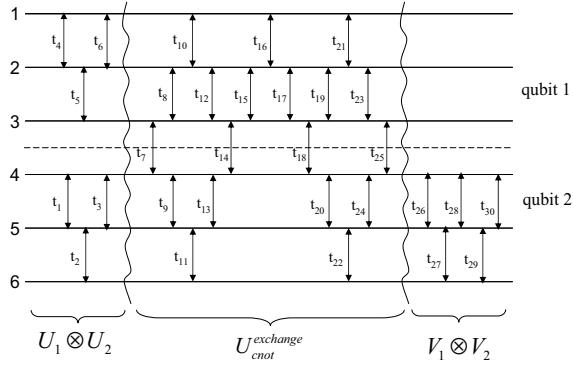}
\caption{Diagrammatic representation of a second, shorter set of
30 exchanges that transform the 19 exchange-only CNOT into the exact
CNOT in the computational basis. Here $U_{cnot}^{exchange}$
corresponds to the 19 gate sequence of Ref.~\onlinecite{divincenzo00}. The
remaining gates and gate times, $t_1$-$t_6$ and $t_{26}$-$t_{30}$,
correspond to the local unitaries ($U_1$, $U_2$, $V_1$, and $V_2$)
which were arrived at analytically by decomposing the local unitaries into a
sequence of elementary rotations in $SU(2)$ as summarized in
Fig.~\ref{fig:fig5}, 
and then using the quaternion representation of the corresponding
rotations in $SO(3)$ to find the sequences of exchange gates and the
times, $t_s$,
that generate them (Appendix~\ref{app:quaternions}). The resulting gate
times are listed in Table~\ref{tab:tab2}.}
\label{fig:fig6}
\end{figure}

The second approach to finding an exchange-only representation of
the local unitaries is
analytic solution through matrix manipulations, as follows. 
We first analyze the similarity transformation $S$ that diagonalizes
$U_{cnot}^{exch}$ from Ref.~\onlinecite{divincenzo00} in the 
computational basis, i.e., $S^{\dagger}U_{cnot}^{exchange}S=D$ where
$D$ is a diagonal 4-by-4 matrix. We found that $S$ can be 
expressed as local operations on the logical qubits, namely 
$S=I \otimes \left( \frac{\sqrt{3}}{2}I-\frac{i}{2}\sigma_y \right)$.      
Mapping the $SU(2)$ spin rotations to rotations in $SO(3)$ and using
the quaternion representation for $SO(3)$ rotations (see 
Appendix~\ref{app:quaternions} for details), we find that this
similarity transformation can be realized using only 3
exchange gates. From the diagonal matrix $D$, one can then readily
generate a C-PHASE gate in the computational basis 
by merely performing rotations around the
z-axis. Transformation of the resulting
C-PHASE into the desired CNOT is subsequently realized by acting with 
Hadamard gates on the second 
logical qubit both before and after the resulting C-PHASE.  These elementary
gates are  
summarized in Fig.~\ref{fig:fig5}. Both the $\sigma_z$
rotations and the 
Hadamard gate have analytic exchange-only solutions on the encoded subspace (see
Sec.~\ref{subsec:singlequbitgates} and Fig.~\ref{fig:fig7}). 
We thereby arrive at an alternative, fully analytic solution for an
exchange-only realization of the local unitary transformations into the
computational basis, requiring a total of 33 exchange gates.  These may be
reduced to a total of 30 exchanges by combining the times of any sequential 
exchanges on the same pairs of qubits that occur as a result of
juxtaposition of elementary gates.
Consequently the desired overall transformation can now be completed with
only 11 more exchanges than the underlying 19-exchange
sequence for $U_{cnot}^{exchange}$.  Fig.~\ref{fig:fig6} and 
Table~\ref{tab:tab2} summarize the resulting gate sequence and gate times,
respectively, for the exact CNOT deriving from this analytic solution of the
local transformations.

The analytical sequence of Table~\ref{tab:tab2} result in a maximum
matrix element deviation of $5.5 \times 10^{-6}$ from the true
CNOT. Thus, our analytic solution has similar accuracy as the
numerical solution above.
However, the 30-exchange analytical sequence in
Fig.~\ref{fig:fig6} and Table~\ref{tab:tab2} represents a saving in both total 
number of gates and total time, relative to the 
35-exchange sequence of Fig.~\ref{fig:fig4} and
Table~\ref{tab:tab1}). 
The 30-exchange sequence requires total time $T_{30}=43.373$,  compared with
$T_{35}=54.326$ for the 35-exchange sequence.   
This is an advantage for experimental implementation, since the shorter time
allows for less decoherence.

\subsection{Single Qubit Gates through Exchange}
\label{subsec:singlequbitgates}
By considering the action of the Heisenberg Hamiltonian 
on the encoded subspace, Eq.~\ref{eqn:encoding}, it was found
numerically in Ref.~\onlinecite{divincenzo00} that arbitrary 
single-qubit gates can be performed on the 3-qubit encoding using 4
nearest neighbour exchanges in serial operation mode, or by using 3
exchanges in parallel. 

\begin{figure}[ht]
\centering
\subfigure[The Hadamard gate can be implemented with 3 exchange gates.]{
  \label{fig:fig7a}
  \includegraphics[width=2.75in,height=1.25in]{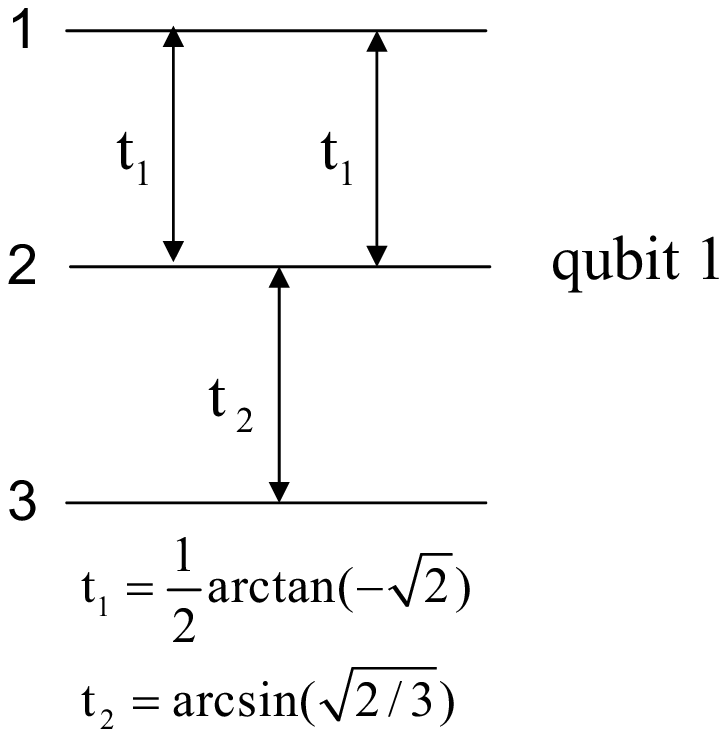}}
\vspace{.2in}
\subfigure[The T gate (T=$\pi/8$ rotation about $\hat{z}$) can be implemented
with just one exchange
gate since $E_{12}=-\sigma_z$, 
where $\sigma_z$ refers to a z-rotation in the encoded
logical subspace.]{
  \label{fig:fig7b}
  \includegraphics[width=2.75in, height=1.25in]{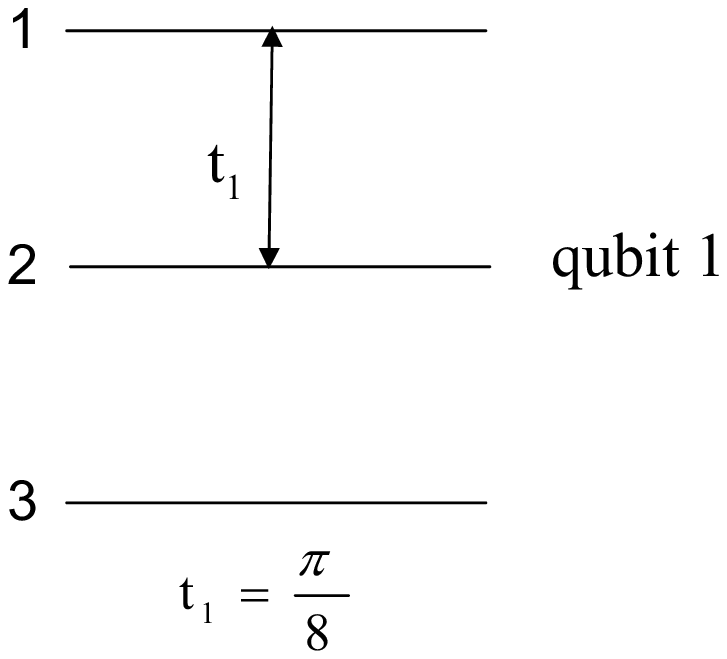}}
\vspace{.2in}
\subfigure[The NOT gate can also be implemented with 3 exchange gates]{
  \label{fig:fig7c}
  \includegraphics[width=2.75in, height=1.25in]{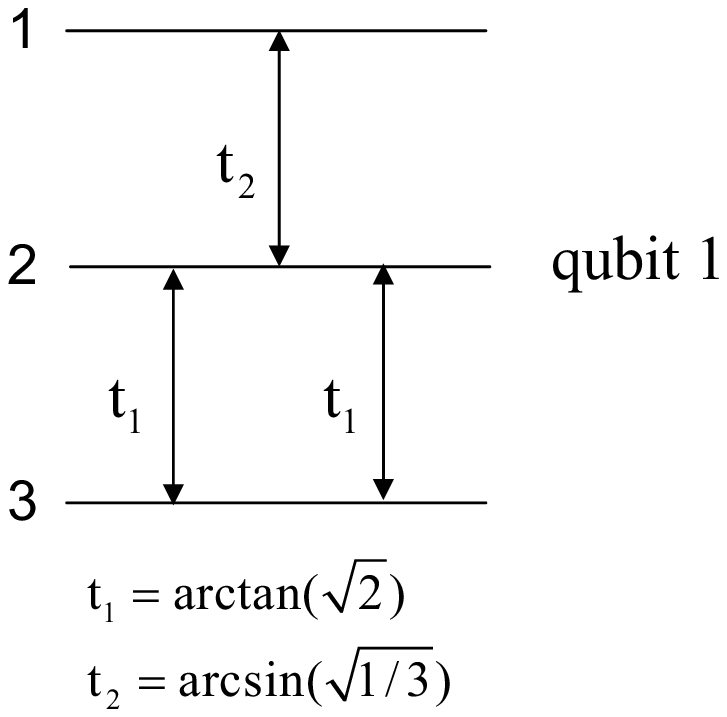}}
\caption{Exchange-only representation of encoded single qubit gates can 
often be arrived at from geometric
considerations. The exchange-only representations of the Hadamard gate, 
the $\pi/8$, and the NOT gate given
here were obtained 
using a quaternion representation for the corresponding rotations in $SO(3)$
(Appendix~\ref{app:quaternions}).}
\label{fig:fig7} 
\end{figure}

The exchange gate times $t_s$ for
a particular single-qubit gate can in principle be found by solving a
system of equations with four unknowns. Specifically, given a
single-qubit gate $A$, we consider the sequence of 4 exchange
gates 
\begin{equation}
A = \exp{( -i t_4 E_{12} / \hbar )} \exp{( -i t_3 E_{23} / \hbar )} \ 
        \exp{( -i t_2 E_{12} / \hbar )} \exp{( -i t_1 E_{23} / \hbar )} 
\end{equation}
and solve the 4 coupled equations for the 4 times $t_s$, $s=1 \,- \, 4$:
\begin{eqnarray}
A_{00} & = &  \exp{( i (t_2+t_4) )} \left( cos(t_3) - \frac{i}{2} sin(t_3) \right) \ 
        \left( cos(t_1)-\frac{i}{2} sin(t_1) \right) \nonumber \\
        & & -\frac{3}{4}\exp{( i (t_4-t_2) )} sin(t_1) sin(t_3) \nonumber \\
A_{01} & = & -i \frac{\sqrt{3}}{2} \exp{( i (t_2+t_4) )} sin(t_1) \
        \left( cos(t_3) - \frac{i}{2} sin(t_3) \right) \nonumber \\ 
        & & -i \frac{\sqrt{3}}{2} \exp{( i (t_4-t_2) )} sin(t_3) \
        \left( cos(t_1) + \frac{i}{2} sin(t_1) \right) \nonumber \\
A_{10} & = & -i \frac{\sqrt{3}}{2} \exp{( i (t_2-t_4) )} sin(t_3) \
        \left( cos(t_1) - \frac{i}{2} sin(t_1) \right) \nonumber \\ 
        & & -i \frac{\sqrt{3}}{2} \exp{(-i (t_4+t_2) )} sin(t_1) \
        \left( cos(t_3) + \frac{i}{2} sin(t_3) \right) \nonumber \\
A_{11} & = & -\frac{3}{4} \exp{( -i(t_2-t_4) )} sin(t_1) sin(t_3)
        \nonumber \\
        & & + \exp{(-i (t_2+t_4) )} \left( cos(t_3) + \frac{i}{2} \
        sin(t_3) \right) \left( cos(t_1) + \frac{i}{2} sin(t_1) \right).
\label{eqn:singlefourunknowns}
\end{eqnarray}

Here we have used the properties of the exchange operators
$E_{12}=-\sigma_z$ and $E_{23}=\sqrt{3}/2\sigma_x+1/2\sigma_z$ in the
logical basis$^($\cite{kempe01,divincenzo00}$^)$.
 
In this work we used a
quaternion approach to represent the single qubit gates as
rotations in $SO(3)$, rather than solving the above coupled
equations. Obtaining the exchange gate times in the quaternion representation also
requires solving trigonometric equations in multiple unknowns, but
these equations, especially for simpler gates, are often more straightforward
than the matrix equation above, 
and analytic solutions easier to obtain.
We first recognize that the combinations of the two exchanges $E_{12}$ and $
E_{23}$ generate the
encoded single qubit operations $\sigma_z$ and $\sigma_x$, and thereby will
suffice to generate any arbitrary rotation on $SU(2)$.  A single qubit gate is 
then mapped from $SU(2)$ to
$SO(3)$$^($\cite{sakurai94}$^)$ where the desired rotation can be
decomposed as a sequence of quaternions. The quaternion approach is convenient
for finding an analytic solution for realization with a given number
of exchanges, as described in Appendix~\ref{app:quaternions}.

Using this approach we found exchange-only gate sequences for the 
$\pi/8$ gate (the T
gate$^($\cite{nielsen00}$^)$), the NOT gate, and the Hadamard gate. A full
description of these solutions is given in
Appendix~\ref{app:quaternions}. We found that both the Hadamard and
the NOT gate can be obtained from a sequence of three exchange gates, while 
the $\pi/8$ gate requires only one exchange gate. The
corresponding gate sequences and 
gate times are shown in Fig.~\ref{fig:fig7}.
(Note that any
$\sigma_z$ rotation by $\theta$ ($R_z(\theta)=exp(-i\theta\sigma_z)$)
can be realized as $exp(i\theta E_{12})$). 

A third approach to finding the exchange-only implementation of the
single qubit gates is through a 
Nelder-Mead simplex numerical
optimization$^($\cite{nelder65,lagarias95}$^)$, as 
implemented for the local transformations in 
the previous section.
Though not analytic, the
Nelder-Mead approach is often much faster than analytic solutions and
for single qubit gates
the cost function can readily be reduced to zero at the machine
precision level. 

\section{Deutsch-Josza Algorithm and Algorithmic Fidelity}
\label{sec:deutschjosza}
There are currently several quantum algorithms that show speed-up
over their classical analogs.$^($\cite{nielsen00}$^)$
Any one of these algorithms serve to investigate the
merits of the exchange-coupled quantum dot implementation.  
\begin{figure}[ht]
\centering
\includegraphics[width=3.5in, height=1.75in]{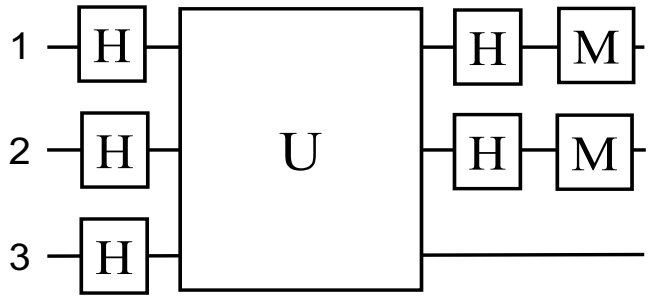}
\caption{A circuit digram depiction of the Deutsch-Josza algorithm for
three qubits. The two top most qubits are the query qubits and make up
the domain for the unknown function $f: \{ 0,1 \} ^n \mapsto \{ 0,1 \}
$$^($\cite{nielsen00}$^)$, here indicated by $U$ a unitary spanning all three
qubits. For the three qubit Deutsch-Josza there are eight
possible U's, all of which are given in
Fig.~\ref{fig:fig9}. Evaluation of $U$ is preceded and
followed by Hadamard gates $[H]$. The result of the function
evaluation is stored in the answer
qubit, the bottom-most qubit in the circuit diagram, as $y \, = \, \ket         {y\; \oplus f(x_1,x_2) \,}$, where $x_1$ and $x_2$ designate the bit
values of query qubit 1 and 2 respectively. The outcome of the
measurement of the query qubits, here designated by M, answers
whether the function is constant or balanced.} 
\label{fig:fig8}
\end{figure}
We have chosen the
Deutsch-Josza algorithm, a relatively simple algorithm requiring only
a few logical gates (Fig.~\ref{fig:fig8}).

\begin{figure}[ht]
\centering
\includegraphics[width=4in, height=1.75in]{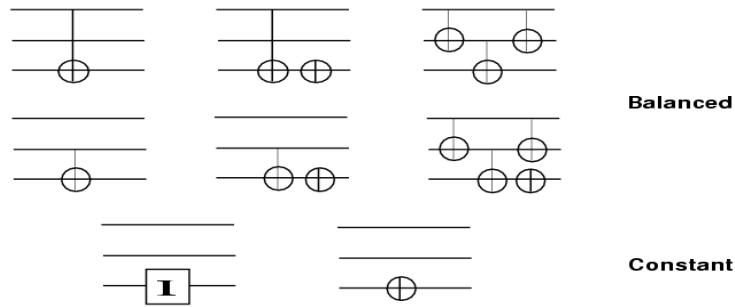}
\caption{Circuit diagram representation of the eight different black
box function evaluations (the U's of Fig.~\ref{fig:fig8})
inherent in the three qubit Deutsch-Josza 
algorithm. For 3 qubits the algorithm results in six balanced
and two constant functions.}
\label{fig:fig9}
\end{figure}

The objective of the Deutsch-Josza algorithm is to determine if an unknown
function, $f: \{ 0,1 \} ^n \mapsto \{ 0,1 \} $,  is balanced ({\em
i.e.} equal number of zeros and ones) or constant. For the
three qubit Deutsch-Josza with its two query qubits and one answer qubit, this
corresponds to 8 possible functions, six 
balanced functions and two constant functions. In
Fig. ~\ref{fig:fig9} these functions are 
expressed in circuit diagram form.

The existence of 8 different functions, and consequently of eight
different versions of the algorithm, implies different
fidelity values for a given initial state, $\rho_0(0)=\rho(0)$, for each
version. For each  
version of the algorithm and for each choice of dephasing rate, we
evaluate the fidelity according to
\begin{equation}
F=Tr\{ \rho_0(t_f) \rho(t_f) \},
\label{eqn:fid}
\end{equation}      
where $\rho_0(t_f)$ is the density matrix describing system evolution
in the absence of errors, and $\rho(t_f)$ the density matrix
describing evolution in the presence of errors. 
Both $\rho_0(t_f)$ and
$\rho(t_f)$ are evaluated over the
time $t_f$ required to complete that version of the algorithm. We
define the 
algorithmic fidelity to be the
worst-case fidelity, namely, for a given value of the dephasing
rate, the fidelity of that version of the
algorithm having the lowest fidelity. This provides a more
conservative estimate than averaging over
the 8 different fidelities resulting from each version of the algorithm.

\section{Numerical Methods}
\label{sec:numericalsimulations}
As discussed in section~\ref{sec:exchange}, universal computation with
the exchange interaction requires at least three physical qubits for every
encoded logical qubit.$^($\cite{kempe01}$^)$ Thus, the Hilbert space
grows as $8^n$, where $n$ is the number of logical qubits, and
simulation of the density matrix becomes 
time consuming even for a few logical qubits. To permit
consideration of larger 
numbers of logical 
qubits we use here the Monte Carlo wave function
method.$^($\cite{dalibard92}$^)$ This 
approach scales linearly with the size of the Hilbert space rather than
quadratically as master equation methods.
  
\subsection{Monte Carlo Wave-functions}
\label{sec:Monte Carlo Wavefunctions}
The Monte Carlo
wave function approach, also known as the method of Quantum
Trajectories or the ``Quantum Jump'' approach, was originally developed within the quantum optics
community.$^($\cite{dalibard92,molmer93,hegerfeldt91}$^)$  The method
relies on a twofold approach to the system evolution. First, an
effective Hamiltonian gives rise to a continuous non-unitary
evolution of the system wave function. Second, decay operators,
identical to the Lindblad operators within the master equation
formalism$^($\cite{carmichael93}$^)$, give rise to stochastic discontinuities in the 
wave function. These stochastic discontinuities resemble the jumps one
might expect from a single isolated quantum system.  

Both the effective Hamiltonian and the decay terms can be
obtained from the 
Lindblad master equation.$^($\cite{carmichael93}$^)$ The conditional, or
effective, Hamiltonian is given by
\begin{equation}
H_{cond} = H_{sys} - \frac{i\hbar}{2} \sum_{m} C_{m}^{\dag}C_{m},  
     \label{Hcond} 
\end{equation}
where $H_{sys}$ is the system Hamiltonian and the $C_m$'s are decay
terms resulting from the
system-environment 
interaction with a subsequent tracing over the bath degrees of freedom. 
The total time evolution under $H_{cond}$ is discretized and at
each time step the probability of any collapse event is calculated:
\begin{equation}
P_{tot} = \sum_{m} \delta P_{m}  = \sum_{m} \Delta t \ 
        \bra{\Psi(t)} C_{m}^{\dag}C_{m} \ket{\Psi(t)}.  \label{Ptot}
\end{equation} 

This total collapse probability accounts for the occurrence of any
error event, $C_m^{\dag}C_m$, that collapses
the system wave-function. The calculated $P_{tot}$ is compared against a random
number, $r$, taken from a uniform distribution. This is the first Monte
Carlo test. A random number less than the total collapse
probability, $r < P_{tot}$, designates that an error has
occurred. Another Monte Carlo 
test, involving another random number, decides which error
occurs. This second random number, $s$, is compared against
the normalized collapse probabilities, $\delta P_{m} / P_{tot}$, and
that error is chosen that first makes the sum of the normalized error
probabilities greater than the random number. Thus, upon completion of
the two Monte Carlo tests the new wave-function is:
\begin{equation}
\ket{\Psi(t + \Delta t)} = \frac{C_i\ket{\Psi(t)}}{ | \bra{\Psi(t)}
C_{m}^{\dag}C_{m} \ket{\Psi(t)} | ^{1/2}},
\label{eqn:collapsewf}
\end{equation}
where $C_i$ is the error operator randomly chosen in the second Monte
Carlo test such that $\sum_{m=1}^{i} \delta P_{m}/P_{tot} > s$.

On the other hand, if $r$ is 
greater than $P_{tot}$, the system state is propagated according to
$H_{cond}$ and we obtain the system state at $t + \Delta t$:
\begin{equation}
\ket{\Psi(t + \Delta t)}' = \exp{\left( -i\Delta t H_{cond} / \hbar \right) }
\ket{\Psi(t)}.
\label{eqn:condwf1}
\end{equation}    
Since $H_{cond}$ is non-Hermitian the norm decreases over
time. To ensure equivalence with other approaches to simulating
open quantum systems, {\em e.g.} master
equations$^($\cite{carmichael93}$^)$, the wave-function 
must be renormalized at the end of every time step: 
\begin{equation}
\ket{\Psi(t + \Delta t)} = \frac{\ket{\Psi(t + \Delta
t)}'}{ | \bra{\Psi(t + \Delta t)}'\ket{\Psi(t + \Delta t)}' | ^{1/2}}.
\label{eqn:condwf2}
\end{equation}
Upon renormalization a new total collapse probability is
calculated and the entire algorithm begins anew for the next timestep.
 
The time step, $\Delta t$,
must be chosen such that $P_{tot} \ll 1$, since for too large time
steps a perturbative expansion for calculating the error probabilities
is no longer justified.  
Each trajectory corresponds to a possible evolution of a single
quantum system. The fidelity measure is based on the density matrix
which can be regained by averaging over
many trajectories.$^($\cite{molmer93}$^)$ Use of Eq.(~\ref{eqn:fid}) leads to 
the following expression for the fidelity:
\begin{equation}
F=\frac{1}{N_{traj}} \sum_{n=1}^{N_{traj}} \mid \bra{\Psi_0(t_f)}
\ket{\Psi_n(t_f)} \mid^2. 
\end{equation}
Here $\ket{\Psi_n(t_f)}$ is the wavefunction for trajectory
$n$ propagated with decoherence, and $\ket{\Psi_0(t_f)}$ is the
wavefunction propagated in the absence of decoherence.
Simulations are run with increasing numbers
of trajectories until the fidelity converges. To
ensure that the fidelity we obtain contains no artifacts or
anomalies due to the choice of the initial system state, we sample a random
distribution of initial states, all located on the surface of the
hyperdimensional Bloch sphere of logical basis states. These Bloch
states are given by: 

\begin{eqnarray}
\ket{\Psi(t)} =  & \sum_{i=0}^{N-1} c_i(t) \ket{\psi} & \nonumber \\
c_k(0) =  & \exp{(i\varphi_k)} \prod_{i=0}^{N-k} & \cos{\theta}_i  \ 
         \prod_{j=N-k}^{N-1} \sin{\theta}_j \nonumber \\
        \varphi_0 = 0, &   & \theta_0 = \pi.  
\label{eqn:blochsphere}
\end{eqnarray}

\subsection{Split Operator Method}
\label{subsec:Split Operator}
The dimensionality of the CNOT gate simulation on 6 physical
qubits, a $2^6=64$ dimensional Hilbert space, is still small enough to permit
the use of an exact diagonalization method to construct the
conditional time evolution operator from $H_{cond}$. However, the three qubit
Deutsch-Josza algorithm requires nine physical qubits within the exchange-only
model, and $H_{cond}$ must then be exponentiated on a $2^9=512$
dimensional space. We have developed a 
more efficient method to construct $U_{cond}=exp(-iH_{cond}t / \hbar )$, that
proves computationally 
efficient for even larger Hilbert spaces. We make use
of a split-operator decomposition of $U_{cond}$ that is based upon the
fact that $H_{cond}$, including decay elements, can be split up into a
term diagonal in the spin components
and a term off-diagonal in the spin components. The diagonal part can
be expressed as  
\begin{equation}
D  = J \, S_{i,z} \cdot S_{j,z} - 2i\hbar\Gamma_{dep}
\sum_{i=1}^{N} S_{z,i}^{\dagger}S_{z,i}, 
\end{equation}
and the off-diagonal part as
\begin{equation}
T  =  \frac{J}{2} (  S^+_iS^-_j + S^+_iS^-_j ) - 2i\hbar\Gamma_{emi}
\sum_{i=1}^{N} S^+_i S^-_i. 
\end{equation}
Here $\Gamma_{dep}$ and $\Gamma_{emi}$ are the single spin pure dephasing
and emission rates obtained from theoretical and experimental
estimates and
measurements~\cite{awschalom99,gupta99,kikkawa97,fujisawa01}.

The time evolution operator, $U_{cond}$, can then be expanded in an approximation
accurate up to second order in $\Delta t$ (errors ~$O(\Delta t^3)$) as:  
\begin{equation}
U_{cond} \approx \exp{ \left( -i D \Delta t / 2 \hbar \right) } \
\exp{ \left( -i T \Delta t / \hbar \right) } \
\exp{ \left( -i D \Delta t / 2 \hbar \right) }. 
\label{splitop}
\end{equation}
The simulation of $U_{cond}(t)$ now reduces to consecutive application
of the exponentiated operators D and T. This can be efficiently done
if states are represented by integers in binary notation, {\em i.e.}
each spin is represented in the $S_z$ basis by a 0 or a 1 at location j
in the binary representation of the state vector $x=\ket{k_1...k_N}$.
The spin operators can be recast as
binary shift and logic operations 
\begin{eqnarray}
S_z \ket{k_1k_2...k_j...k_N} & = & \frac{1}{2}(1-2*ibits(k,j,1)) \, \ket{k_1k_2...k_j...k_N}, \\
S^+ \ket{k_1k_2...k_j...k_N} & = & (1-ibits(k,j,1)) \ket{k_1k_2...ibits(k,j,1)+1...k_N}, \\
S^- \ket{k_1k_2...k_j...k_N} & = & ibits(k,j,1) \ket{k_1k_2...ibits(k,j,1)-1...k_N},
\end{eqnarray}
where {\em ibits} denotes a compiler (F90) command that extracts
the value of jth bit in integer k. These binary operations are seen to
act upon the system state vector in a manner analogous to raising and lowering operators. With this
approach it becomes possible to simulate very large Hilbert
spaces. This approach was employed previously in a checkerboard time
propagation scheme for 
study of many body dynamics of interacting particles on
lattices.$^($\cite{zhang91a,zhang92}$^)$ 

\subsection{Parameters}
\label{subsec:parameters}
Data with regards to decoherence
parameters for exchange coupled quantum dots is scarce. We have used experimental parameters to the extent
possible. Where none are available we have interpolated, using
theoretical estimates, between what is experimentally known
and the requirements of our simulations.
In general, experiments in condensed matter physics have indicated
that the electron spin states, because of their weaker coupling to the
environment, exhibit longer coherence times than the charge
states. Due to difficulties involved in measuring single spin states,
however, the majority of these experiments provide us with a ensemble
measurement of the lifetime, and are thus not directly
applicable to a system of single
spins$^($\cite{awschalom99,gupta99,kikkawa97}$^)$ We employ here
the inequality relationship $T_1 \geq T_2 \geq T_2^*$, where $T_1$
describes the time scale for the spins' exchange of energy with the
surrounding matrix,
$T_2$ is the single spin decoherence time, and $T_2^*$ is an ensemble
decoherence time which, in addition to contributions from $T_1$ and $T_2$,
also contains effects due to inhomogeneities in the system, to the
surrounding matrix, and to the control fields.$^($\cite{engel01}$^)$ Taking
into account the single dot $T_1$ times obtained
by Fujisawawa {\em et al.}$^($\cite{fujisawa01}$^)$, we arrive at a set of reasonable
decoherence parameters: a dephasing rate on the order of ns and a
timescale of $\mu$s for emission and absorption, both of which involve
spin flips. Consequently, in a system like
ours, where the strength of the exchange coupling is assumed to be on
the order of 0.2 $meV$, we find dimensionless decoherence rates
$\hbar \Gamma / J_0 \, = \, 10^{-3} \, - \, 10^{-5}$. Additionally, we find
that dephasing errors dominate over emission events, according to    
$\frac{\Gamma_{dep}}{\Gamma_{emi}}=\frac{1/T_2}{1/T_1} \simeq
10^2$.  
Consequently, pure dephasing is a greater concern than spin flip errors,
and will thus constitute 
the main focus of our simulations. In the context of decoherence,
it should be noted that a reduction of some of
the decoherence pathways may be possible with the use of experimental
techniques such as spin polarization and spin echo, that have been
developed for other systems such as NMR.$^($\cite{hahn50}$^)$ Our simulation
does not include these
potentially very beneficial modifications. Note that the encoding in
Eq.~(\ref{eqn:encoding}) is automatically protected against collective
dephasing, but not against independent single spin dephasing.

\section{Results}
\label{sec:results}
\subsection{Exchange-only CNOT, in serial mode}
\label{subsec:resultcnot}
\begin{figure}[ht]
\centering
\includegraphics[width=4in, height=3in]{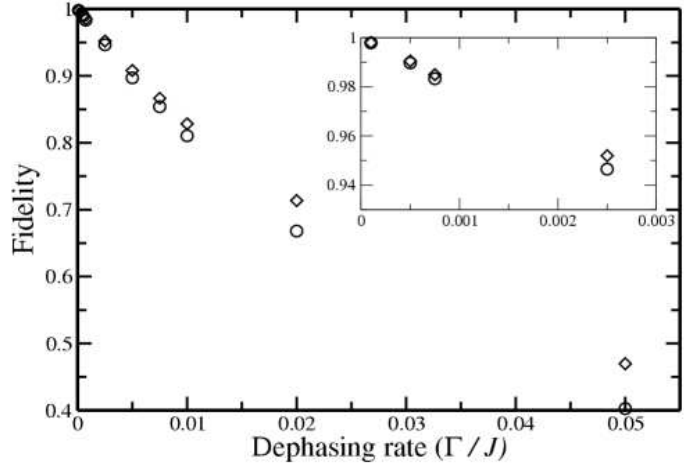}
\caption{Fidelity simulation for CNOT gate, subject to
dephasing errors, for a linear quantum dot array, where three
physical qubits encodes each logical qubit. $\circ$ 19 
exchange gate implementation of CNOT gate. $\diamond$ Free
system evolution for duration equivalent to CNOT.}
\label{fig:fig10}
\end{figure}

The encoded exchange-only CNOT is the first unitary operation
we investigate. Fig.~\ref{fig:fig10} shows the 
fidelity over the 19 gate implementation. We see that for a
dimensionless dephasing rate of $\hbar \Gamma \, / \, J_0 \, = \, 10^{-3}$ the
probability of perfectly 
performing a CNOT is $ \sim 98 \%$. 

\begin{figure}[ht]
\centering
\includegraphics[width=4in, height=3in]{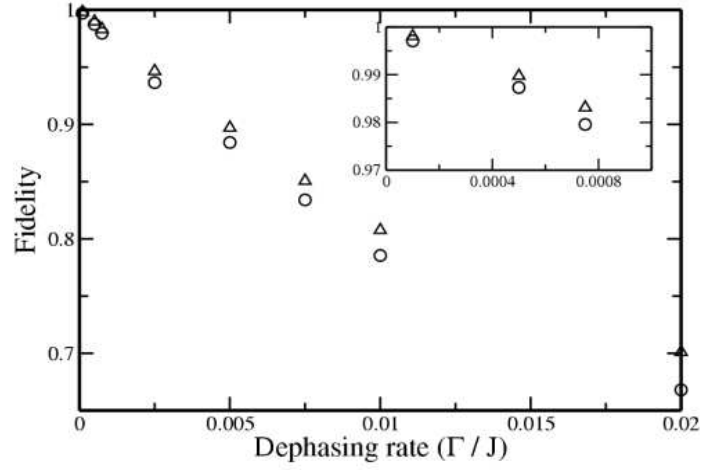}
\caption{Fidelity simulation for a CNOT gate, subject to
dephasing errors, for an encoded linear quantum dot array and
between two physical qubits. $\circ$ 19 exchange gate
implementation of CNOT gate. $\bigtriangleup$ CNOT gate implemented
between two physical qubits, with gate time equal to 19
exchange gates.}
\label{fig:fig11}
\end{figure}

Burkard {\em et al.} have indicated a possibility that actual
implementation of the 
gate, {\em i.e.} turning on the $J_0$ coupling between adjacent
quantum dots, might result in faster
decoherence.$^($\cite{burkard99}$^)$ We therefore used the free system evolution under identical
conditions of dephasing as a point of reference. We find that gate
implementation does result in faster decoherence, 
but that the effects only become appreciable at higher dephasing
rates, $\hbar \Gamma_{dep} / J_0 = 10^{-2}$ (Fig.~\ref{fig:fig10}). We also 
compared the fidelity obtained for the encoded CNOT gate to a standard
CNOT gate between two physical qubits. To reduce the effects of method
and parameter choice, we used the same values for the dephasing rate
to interdot coupling strength ratio, $\hbar \Gamma_{dep}/J_0$, and
took the timescale for the 
CNOT gate to be the same as for the exchange coupled
qubits. This comparison is shown in Fig.~\ref{fig:fig11}. 

Fig.~\ref{fig:fig11} shows that the
performance of the encoded CNOT gate deteriorates faster with
increasing dephasing rates than does the bare CNOT gate. We recall
that the timescale for coherence loss due to dephasing 
decreases with the number of qubits. As shown in Ref.~\onlinecite{carmichael93} the
master equation for a single qubit under pure dephasing
can be written as: 
\begin{equation}
\dot{\rho}(t)=-\frac{i\omega}{2} [\sigma_z,\rho(t)]+\frac{\Gamma_{dep}}{2}
(\sigma_z\rho(t) \sigma_z - \rho(t)),
\label{eqn:2leveldep}
\end{equation}
where $\omega$ is the shifted frequency of the two level system and
$\Gamma_{dep}$ is the dephasing rate. The dynamics of
Eq. (\ref{eqn:2leveldep} can be solved analytically and for this
2-level system the fidelity (see Eq. (\ref{eqn:fid})) is obtained as:
\begin{eqnarray}
F & = & Tr \, \left\{ \left( \begin{array}{cc}
\rho_{00}(0) & \rho_{01}(0)e^{-iwt} \\
\rho_{10}(0)e^{iwt} & \rho_{11}(0) \end{array}
\right) \left( \begin{array}{cc}
\rho_{00}(0) & \rho_{01}(0)e^{-iwt-\Gamma_{dep}t} \\
\rho_{10}(0)e^{iwt-\Gamma_{dep}t} & \rho_{11}(0) \end{array}
\right) \right\} \nonumber \\
& = & \rho_{00}^2(0)+\rho_{11}^2(0)+e^{-\Gamma_{dep}t}(|\rho_{01}(0)|^2+|\rho_{10}(0)|^2)
\end{eqnarray}
This state-dependent fidelity must now be integrated over all possible
initial states to obtain the algorithmic fidelity. Employing a
general state
$\Psi(0)=\cos{\frac{\theta}{2}}\ket{0}+e^{i\phi}\sin{\frac{\theta}{2}}\ket{1}$
and integrating over all possible states on the surface of the Bloch
sphere leads to the average fidelity: 
\begin{eqnarray}
\label{eqn:fiddep}
\bar{F}(t) & = & \frac{1}{4\pi} \int_0^{2\pi} \int_0^{\pi}
Tr \{ \rho_0(t)\rho(t) \} \sin(\theta) d\theta d\phi \\
& = & \frac{2+e^{-\Gamma_{dep}t}}{3}, \nonumber   
\end{eqnarray}
This average fidelity asymptotically approaches
$2/3$ for a single qubit and $(\frac{2}{3})^N$ for N independent qubits as
$t \, \rightarrow \, \infty$. Now the CNOT 
gate involves couplings between qubits, and hence the long-time CNOT
gate fidelity dependence upon the number of qubits will not be
exactly $(\frac{2}{3})^N$. However the overall faster decay of the
fidelity as $N$ increases is still found. This is also evident in
Fig.~\ref{fig:fig11}.  

\begin{figure}[ht]
\centering
\includegraphics[width=4in, height=3in]{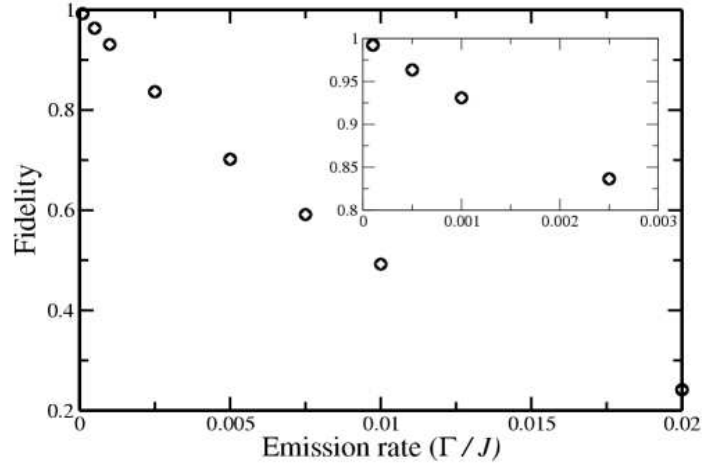}
\caption{Fidelity simulation for CNOT gate, subject to emission
errors, for a linear quantum dot array, where three physical qubits
encodes each logical qubit. $\circ$ Nineteen exchange gate implementation
of CNOT gate. $\diamond$ Free system evolution for duration equivalent to
CNOT.}
\label{fig:fig12}
\end{figure}

The effects of
emission upon the CNOT gate fidelity are summarized in
Fig.~\ref{fig:fig12}. We see here a greater degeneration as a
function of the emission error rate. Emission
events are intrinsically more detrimental to the proper operation of
our quantum 
device than are dephasing errors. They signify a change in the
system's overall 
energy. In contrast, dephasing errors
merely introduce a random phase difference between the ground and
excited states. Note, under the independent error model used here
the logical basis states do not lie in a decoherence free subspace, which
would have been the case for a collective error model.$^($\cite{bacon99} Thus,
both emission errors and dephasing errors will take the system outside
the encoded subspace ($S_z^{tot}=1/2$). We recall that
emission events are generally a much rarer occurrence than dephasing events. As
mentioned in section ~\ref{subsec:parameters}, the expected ratio of
dephasing to emission in semiconductor quantum dots is
$\frac{\Gamma_{dep}}{\Gamma_{emi}}=\frac{1/T_2}{1/T_1} \simeq 
10^2 \: - \: 10^3$. In contrast, for $\hbar \Gamma_{dep}/J_0 \sim 10^{-3}$ we have
$\hbar \Gamma_{emi}/J_0 \sim 10^{-5}-10^{-6}$. As seen in
Fig.~\ref{fig:fig12} (inset), in this regime the encoded gate
fidelity is $\geq \, 95 \, \%$.

\subsection{Algorithmic Fidelity of Three qubit Deutsch-Josza}
\label{subsec:resultdj}
For the the three qubit Deutsch-Josza algorithm there are eight
possible function evaluations, listed in Fig.~\ref{fig:fig9}
(Sec.~\ref{sec:deutschjosza}). 
\begin{figure}[ht]
\centering
\includegraphics[width=4in, height=3in]{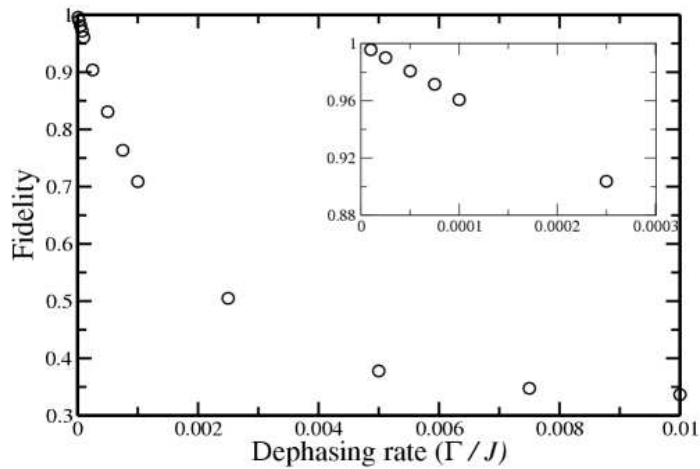}
\caption{Simulation of three qubit Deutsch-Josza algorithm,
where the fidelity plotted is the worst case fidelity, {\em i.e} the
lowest calculated fidelity for any of the eight possible
functions.} 
\label{fig:fig13}
\end{figure} 
We found an algorithmic fidelity $F \: \geq \: 0.70$ for dephasing rates
$\hbar \Gamma_{dep} / J_0 \: \leq \: 10^{-3}$ (increasing to $F \, \geq \, 0.98$
for $\hbar \Gamma_{dep} / J_0 \: \leq \: 10^{-5}$). The algorithmic fidelity is
shown as a function of $\hbar \Gamma_{dep} / J_0$ in
Fig.~\ref{fig:fig13}.  

\subsection{Using N-qubit unitaries to simplify algorithm implementation}
\label{subsec:algorithmicreduction}
From the perspective of quantum computing, another approach 
to overall time reduction is possible. As noted in
Refs.~\onlinecite{sanders99} and \onlinecite{niskanen03}, any sequence
of logical gates may be replaced by a single N-qubit unitary. Thus, in
Sec.~\ref{sec:exchange} 
it was shown that the sequence of four adjacent qubit CNOTs is equivalent to
CNOT(1,3).We may similarly reduce combinations of other single and
two qubit-gates to just one N-qubit gate involving only 2 body
interactions. Certainly, many of these N-qubit gates  
will not be as simple as the CNOT(1,3). They may nevertheless allow
for a faster, more efficient experimental realization of certain
combinations of gates.

\begin{figure}[ht]
\centering
\includegraphics[width=3in, height=1.75in]{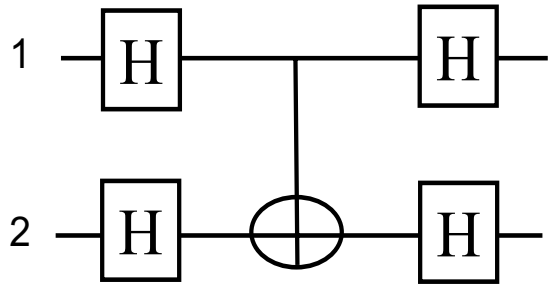}
\caption{The 2-qubit unitary consisting of a CNOT
sandwiched between 4 Hadamards.}
\label{fig:fig14}
\end{figure}

\begin{figure}[ht]
\centering
\subfigure[Adding the exchange-only representation of the Hadamard gates onto the exact CNOT
  sequence of Fig.~\ref{fig:fig6} results in a 45 gate exchange sequence
that implements the 2-qubit unitary of Fig.~\ref{fig:fig14}.
  This sequence can be reduced by adding the 
  gate times for consecutive exchanges acting on the same qubit pair 
  and by making use of the relation $H^2 = I$ for the Hadamard gate to
arrive at sequence b) below.]{
  \label{fig:fig15a}
  \includegraphics[width=4in,height=1.75in]{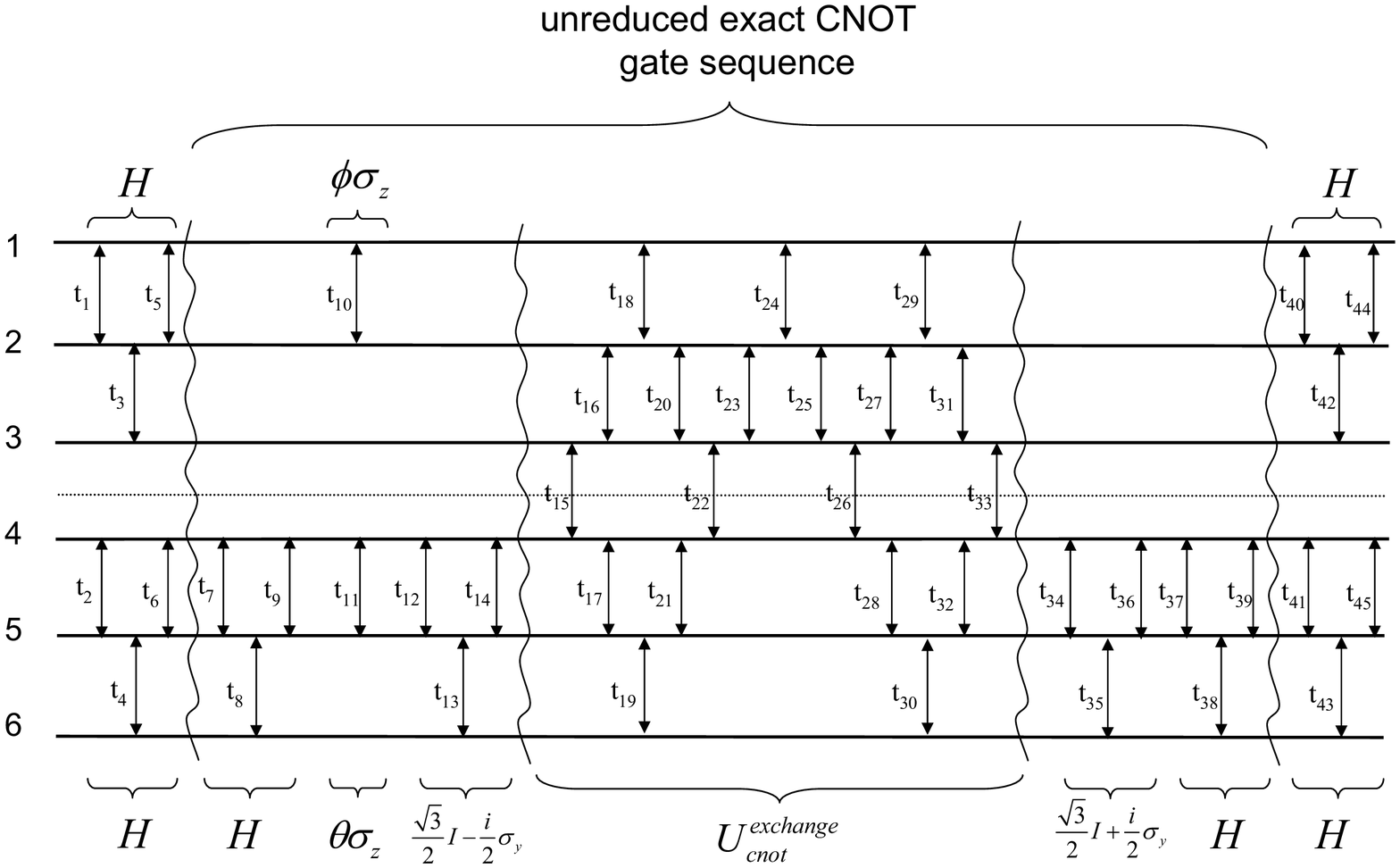}}
\vspace{.2in}
\subfigure[Resulting shorter sequence of 31 exchange gates that directly implements the
2-qubit unitary of Fig.~\ref{fig:fig14}. 
The corresponding gate times are given in
Table~\ref{tab:tab3}.]{
  \label{fig:fig15b}
  \includegraphics[width=4in,height=1.75in]{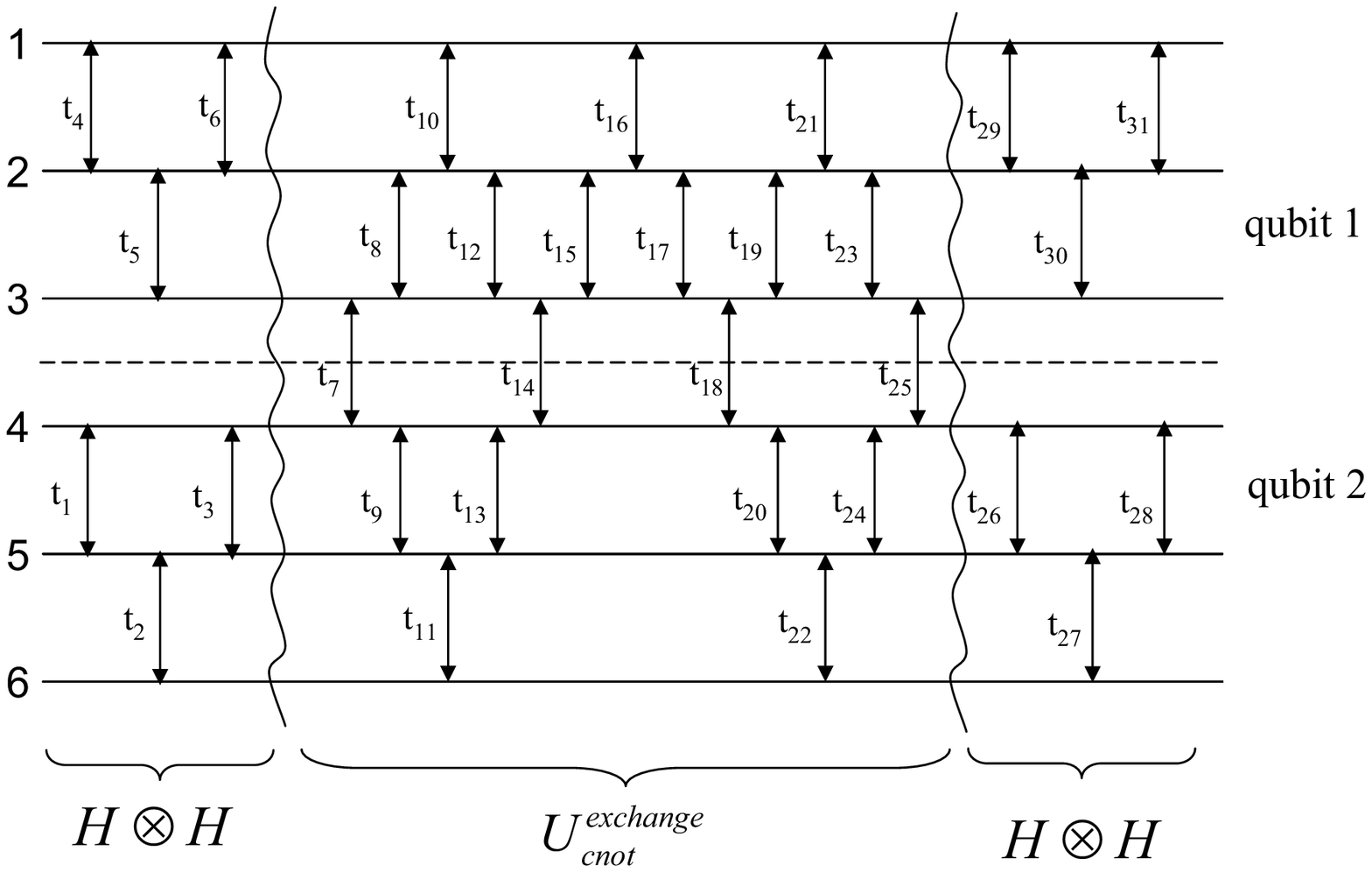}}
\caption{}
\label{fig:fig15} 
\end{figure}

We have analyzed this approach for the example of a CNOT sandwiched
between 4 Hadamard gates (Fig.~\ref{fig:fig14}). This circuit, which can 
readily be verified to be 
equivalent to a CNOT with the control and target qubits reversed, 
constitutes a relatively simple two-qubit unitary where it is possible
to find an 
analytic solution for exchange-only implementation,
using the quaternion decomposition described 
in Appendix~\ref{app:quaternions}. Starting with the original unreduced version
($3+1+1+3+19+3+3=33$ exchange gates) of the exact analytic CNOT
sequence shown in Fig.~\ref{fig:fig6}, the exchange gates
corresponding to the Hadamard gates are then added before and after this
sequence to arrive at a 45 exchange gate sequence for the desired
2-qubit unitary, shown in Fig.~\ref{fig:fig15a}).  The length of this sequence can be reduced 
by using the relation $H^2 =I$ for the Hadamard gate and by combining the
gate times for consecutive exchange gates acting on the same qubit pair, as
was done for Fig.~\ref{fig:fig6}.  This yields
the 31 exchange gate long sequence shown in 
Fig.~\ref{fig:fig15b}. The corresponding exchange gate times are
listed in Table~\ref{tab:tab3}.  
In more general cases of $N$-qubit unitaries involving 
many consecutive two-qubit and one-qubit gates, this analytic approach might not be 
feasible
and numerical optimization techniques such as that described in
Section~\ref{subsec:exactcnot} will then have to be used. 

The sequence of gates shown in Fig.~\ref{fig:fig15} contains 31
gates. This analytic solution should
be compared against the 42 gates required for implementing the gates
consecutively using $4\, \times \,3$ exchange gates to represent the
four Hadamard gates and 30 exchange gates to represent the
CNOT. One expects that with fewer gates and shorter
total implementation time, better fidelities would 
result. 
\begin{figure}[ht]
\centering
\includegraphics[width=4in, height=3in]{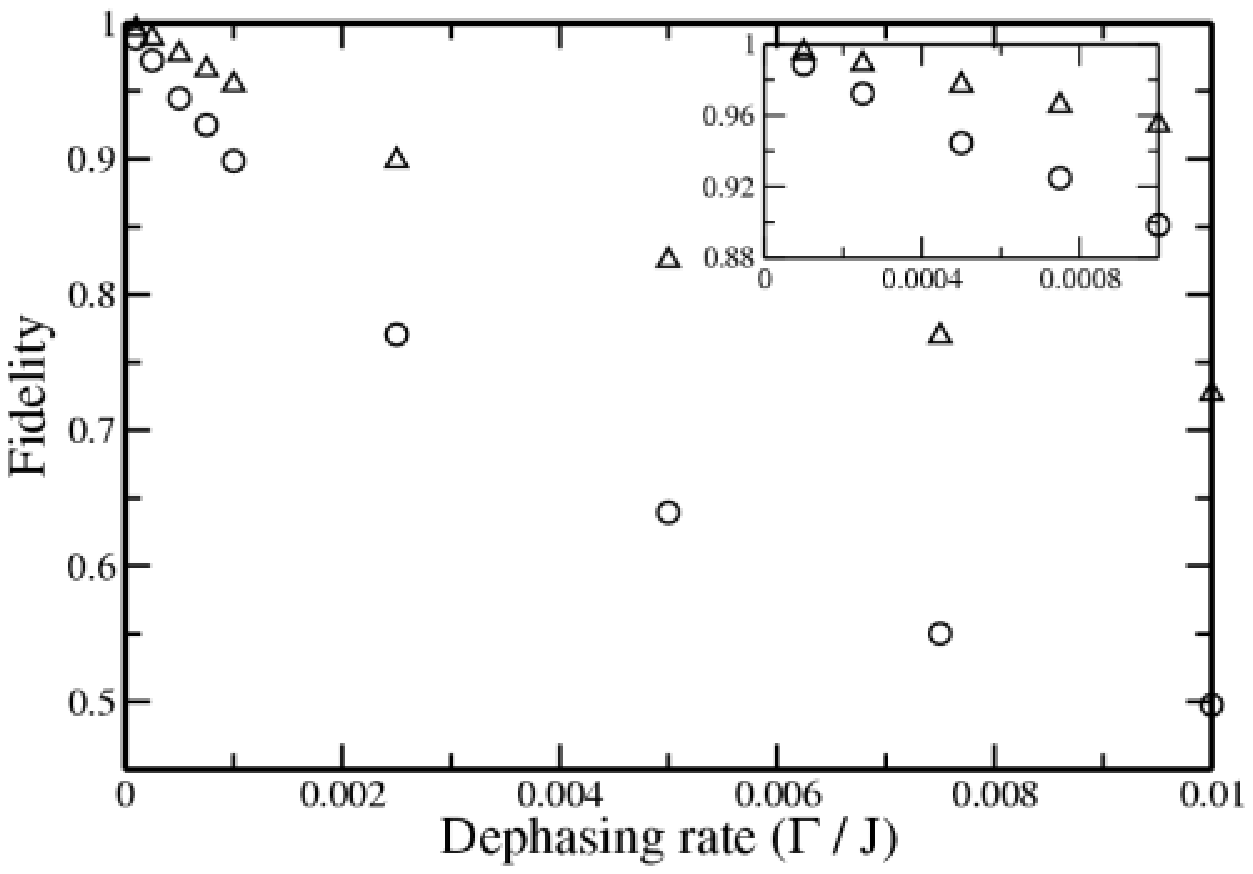}
\caption{Comparison of fidelity obtained using a standard serial
implementation of the logical gates with 42 exchange operations, versus
the 31 gate sequence shown in Fig.~\ref{fig:fig15}. The simulations employed
20 time steps per exchange gate and 25600 trajectories averaged over at
least 64 input states sampled from the surface of the hyperdimensional
logical Bloch sphere, Eq.~\ref{eqn:blochsphere}. 
Open circles ($\circ$): the normal serial
implementation of the logical gates in Fig.~\ref{fig:fig14} (42
exchanges).  Open triangles 
($\bigtriangleup$): the overall unitary gate is decomposed into
individual exchange gates according to Fig.~\ref{fig:fig15} (31 exchanges).}
\label{fig:fig16}
\end{figure}
Fig.~\ref{fig:fig16} shows that this is indeed the case. The
fidelity for the shorter sequence is about 5 
to 10 percent better at decoherence rates ($\hbar \Gamma_{dep} / J_0$) $\sim
\, 10^{-3}$, and the
improvement is even greater for faster decoherence rates.

\section{Summary and conclusions}
\label{sec:conclusions}
We have investigated here the merits of exchange-only quantum
computation based on a linear quantum dot array. For this
architecture, we have shown that it is possible to achieve fidelities
of 95 $\%$ or greater for the CNOT gate with realistic choices of
system parameters.  We
have also elucidated the performance under dephasing of the 3 qubit
Deutsch-Josza algorithm that is implemented with a 3 qubit encoding. For this
algorithm we obtained a fidelity of at least 0.70 for realistic
dephasing rates, $\hbar \Gamma_{dep} / J_0 \leq 10^{-3}$. In addition, we have
provided an example and supporting simulation data for replacing a series of
gates with a single unitary. This approach is advantageous because
it reduces the total time for implementing the algorithm, and will
thus thereby also reduce the effects of decoherence.

Our results indicate that, due to the currently rather high decoherence rates, achieving the $10^{-4}$ threshold
required for fault tolerance is beyond present capabilities in this
spin-coupled quantum 
dot model. Nevertheless, the success probabilities for the
Deutsch-Josza simulations imply that exchange coupled quantum dot
arrays make for an interesting 
testbed. With improved experimental solid state technology,
greater gate fidelities can be expected. Getting to $\hbar \Gamma_{dep} / J_0
\, = \, 10^{-5}$ will yield $\geq 98 \, \%$ algorithmic fidelity
(Fig.~\ref{fig:fig13}). 
In the context of extending the relevance of these simulations,
and how they pertain to other systems, it should be noted that a perfectly
isotropic interaction is not a necessity for universality, as it has
recently been shown that both the anisotropic and asymmetric
interactions are universal under appropriate encoding.$^($\cite{vala02}$^)$ 

We have attempted to provide here a realistic estimate of gate and
algorithmic fidelity for exchange-only quantum computation. Our estimates could be improved by
having more realistic single spin parameters and by incorporating pulse
shaping techniques. The square
pulses assumed here provide only an approximation to experimental
pulses. However, since the ability to
implement $SU(4)$ and $SU(2)$ operations is only dependent on
integrated pulse shape, square pulses are adequate from a theoretical
perspective, provided that the qubit is defined on a pure two-level system
and a square
pulse therefore cannot cause excitation to higher levels. In
the future it would be desirable to perform simulations where the pulses better
reflect what is achievable in the laboratory. Allowing for
pulse shaping and employing chirped pulses have been shown to improve
both gate and 
algorithmic fidelities$^($\cite{chen01a}$^)$, making such simulations doubly
interesting for 
future work.      

\clearpage

\section{Acknowledgments}
We thank David Bacon, Kenneth Brown, and Patrick Huang for
useful discussions. The effort of the
authors is sponsored by the Defense 
Advanced Research Projects Agency (DARPA) and the Air Force
Laboratory, Air Force Material Command, USAF, under Contract
No. F30602-01-2-0524, and DARPA and the Office of Naval Research under
grant No. FDN 00014-01-1-0826. Additional support was provided by the
National Security Agency under contracts DAAD 19-00-1-0380 and DAAG
55-98-1-0371. We also
thank NPACI for a generous allocation of supercomputer time at the San Diego
Supercomputer Center.

\begin{appendix}
\section{Quaternions}
\label{app:quaternions}
First developed by 
Hamilton, quaternions provide an alternative to the normal matrix
representation of vectors and rotations in $SO(3)$.$^($\cite{hamilton67,kuipers98}$^)$   

\begin{figure}
\centering
\includegraphics[width=3in, height=2.5in]{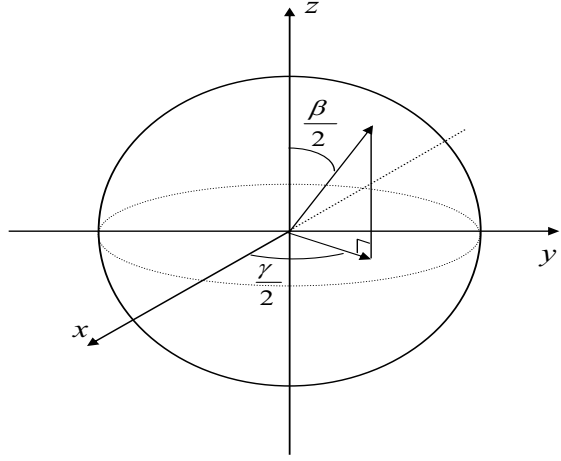}
\caption{The angles $\beta$ and $\gamma$ are used to define the axis
  of rotation when using hyperspherical coordinates.}
\label{fig:fig17}
\end{figure}

A general quaternion,
$q=\{w,q_x,q_y,q_z\}=\{w,\underline{u}\}$, is just a four component
array which can 
represent either a vector in $\Re^3$ or an $SO(3)$ rotation. The first
component represents a scalar and the last three components a vector. More specifically,
vectors and rotations take the form:
\begin{eqnarray}
\label{eqn:quaternions}
v & = & \{ 0, \, v_x, \, v_y, \, v_z \} \nonumber \\
q_r & =  & \{ cos(\alpha/2), \; r_xsin(\alpha/2), \; r_ysin(\alpha/2),
\; r_zsin(\theta/2) \}
\end{eqnarray}
Here $q_r$ corresponds to a rotation of $\alpha$ around the Cartesian
vector $\hat{r}=(r_x, \, r_y, \, r_z)$. Using hyperspherical coordinates this
vector, and the resulting quaternion, can be written as
\begin{eqnarray}
\hat{r} & = &
(sin(\beta/2)cos(\gamma/2), \, sin(\beta/2)sin(\gamma/2), \, cos(\beta/2))
\nonumber \\
q_r & =  & \{ cos(\alpha/2), \, sin(\alpha/2)sin(\beta/2)cos(\gamma/2),  \nonumber \\
 & & sin(\alpha/2)sin(\beta/2)sin(\gamma/2), \,
sin(\alpha/2)cos(\beta/2)  \},
\label{eqn:quaternionhyper}
\end{eqnarray}
where $\beta$ and $\gamma$ define the axis of rotation, as shown in
figure~\ref{fig:fig17}, and $\alpha$ is the angle of rotation around this axis.
Two quaternions are multiplied together to form a new quaternion: 
\begin{eqnarray}
q_1*q_2 & = & \{w_1, \underline{u}_1 \}*\{ w_2, \underline{u}_2 \}
\nonumber \\
        & = & \{w_1w_2-\underline{u}_1 \cdot \underline{u}_2,
        w_1\underline{u}_2+w_2\underline{u}_1+\underline{u}_1 \times
        \underline{u}_2\}.
\end{eqnarray}
Hence one can derive expressions for sequences of
rotations, $Q=q_N*q_{N-1}*...*q_2*q_1$. For example the sequence of
Euler angle rotations$^($\cite{zare88}$^)$ 
$R(\phi,\theta,\chi)=e^{(-iS_z\phi)}e^{(-iS_y\theta)}e^{(-iS_z\chi)}$
becomes in the quaternion representation:
\begin{eqnarray}
q(\phi,\theta,\chi) & = & \{ cos(\frac{\theta}{2})cos(\frac{\phi}{2}+\frac{\chi}{2}), \,
sin(\frac{\theta}{2})sin(\frac{\chi}{2}-\frac{\phi}{2}), \nonumber \\ 
 & & sin(\frac{\theta}{2})cos(\frac{\chi}{2}-\frac{\phi}{2}),
\,cos(\frac{\theta}{2})sin(\frac{\chi}{2}+\frac{\phi}{2}) \}.   
\label{eqn:euler}
\end{eqnarray}
The Heisenberg Hamiltonian provides us with two rotations on the encoded
qubit, $\sigma_z$ and $\sigma_x$:$^($\cite{kempe01}$^)$
\begin{eqnarray}
e^{itE_{12}} & = & e^{-it\sigma_z} \nonumber \\
e^{itE_{23}} & = & \
e^{it(\frac{\sqrt{3}}{2} \sigma_x + \frac{1}{2} \sigma_z )}.
\label{eqn:exchangegates}
\end{eqnarray}
The first is a rotation around $\hat{z}$, the second a rotation around
the axis $\hat{k}=\sqrt{3}/2 \, \hat{x}+1/2 \, \hat{z}$. Note, as seen
from Eq.~\ref{eqn:heisenbergtoexchange}, using $E_{ij}$ rather than
$H_{ij}$ to calculate gate sequences and gate times just results in a
global phase in the final state, which can be accounted for as follows: 
\begin{equation}
\prod_{k=1}^{M} \exp{( it'_kE_{i_kj_k} )} = \exp{( iJ_0/4\hbar\sum_{k=1}^{M} t_k )}
\prod_{k=1}^{M} \exp{( it_kH_{i_kj_k}/\hbar)}.
\end{equation} 
Where $t'_k=\frac{J_0}{2\hbar}t_k$ results from the rescaling
necessary when using $E_{ij}$ instead of $H_{ij}$.

\begin{figure}[ht]
\centering
\includegraphics[width=3in, height=2.5in]{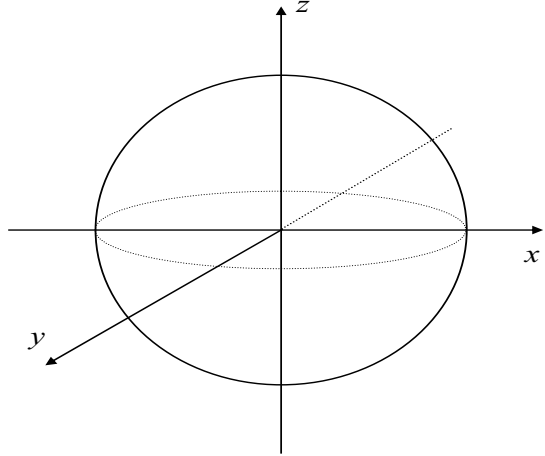}
\caption{The Bloch sphere representation of a spin $1/2$
particle.$^($\cite{sakurai94}$^)$}
\label{fig:fig18}
\end{figure}

Given the
mapping of $SU(2)$ to $SO(3)$ (the Bloch sphere representation for
spin-1/2 systems, see Fig.~\ref{fig:fig18})$^($\cite{sakurai94}$^)$,
these exchange gates (Eq.~\ref{eqn:exchangegates}) can be
cast in the quaternion representation respectively as
\begin{eqnarray}
q_1(t) & = & \{ cos(t/2), \, 0, \, 0, \, -sin(t/2) \}, \nonumber \\
q_2(t) & = & \{ cos(t/2), \, \frac{\sqrt{3}}{2}sin(t/2), \, 0, \, \ 
\frac{1}{2}sin(t/2) \}.
\end{eqnarray}

Having obtained the quaternions that correspond to the different
possible exchange gates on our encoding, and understanding how these quaternions can be
multiplied together, we can now investigate the number of exchanges required to
generate certain single qubit gates. 

\begin{figure}[ht]
\centering
\includegraphics[width=3in, height=2.5in]{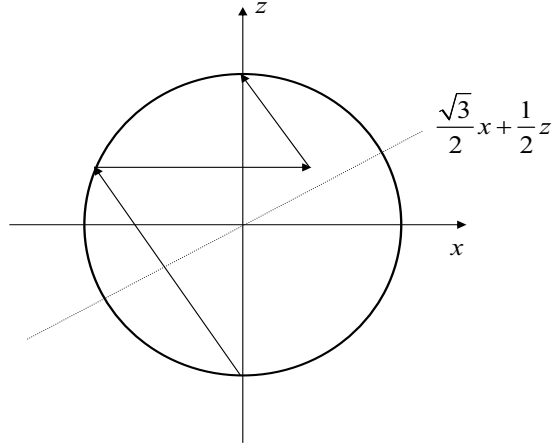}
\caption{Planar projection of the Bloch sphere onto the $x-z$ plane,
depicting the rotations corresponding to $\exp{( itE_{12} )}$ and $\exp{(
itE_{23} )}$. Implementing $E_{12}$ is the
same as rotating around the $\hat{z}$ axis, the action of which causes
states to follow a line parallel to the $\hat{x}$ axis. The $E_{23}$ gate
corresponds to rotations around the axis$\sqrt{3}/2 \, \hat{x}+1/2 \,
\hat{z}$, and states then follow a line
perpendicular to this axis during the duration of the gate. Combined
in the sequence $\exp{(it_3E_{23})}\exp{( it_2E_{12} )}
\exp{( it_1E_{23})}$ these two exchange gates allows us to
reach any point on the Bloch sphere starting from any arbitrary
point. The figure shows how this sequence of gates allows us
to rotate a state from the south pole to the north pole, {\em i.e.} to
the most distant state. The ability to generate all other rotations
follows directly from this.$^($\cite{junzhang03}$^)$ Note, once
at the north pole
it might be necessary to adjust the phase of the state. Thus we
need to add one more $E_{12}$ gate to the sequence, arriving at a
total of at most 4
exchange gates to generate any $SU(2)$ operation. (Note that this
final $E_{12}$ gate serves exactly the same purpose as the final
$\sigma_z$ rotation does in the Euler construction.) }
\label{fig:fig19}
\end{figure}

Using the geometric representation for generation of the $SU(2)$
group of rotations$^($\cite{dalessandro01,junzhang03}$^)$ we investigate
how many sequential 
implementations of our exchange gates, $\exp{( itE_{12} )}$ and $\exp{(
itE_{23} )}$, suffice to generate all possible single qubit
rotations. Projected onto the $x-z$ plane of the Bloch sphere, the
rotations around $\hat{z}$ and $\sqrt{3}/2 \, \hat{x}+1/2 \,
\hat{z}$ corresponding to these exchange gates can be
represented as in Fig.~\ref{fig:fig19}.

Based on this
geometric interpretation for the action of the exchange gates, we can
proceed as in
Ref.~\onlinecite{junzhang03} and determine the minimum number of exchange
gates required to generate any $SU(2)$ rotation from a given pair of
exchanges. To do this we
consider how many exchange gates, and in what order they should be
arranged, suffice to rotate a state from the south pole all the
way to the north pole. As shown in Fig.~\ref{fig:fig19}, this extreme
rotation can be
achieved using 3 exchange gates in the sequence $\exp{(
it_3E_{23})}\exp{( it_2E_{12} )} \exp{( it_1E_{23}
)}$.  This rotation from one pole to the other is the hardest to
achieve, in the sense that it requires the most changes of direction
and hence the greatest number of exchanges. All other rotations will
require equal or less exchanges.$^($\cite{junzhang03}$^)$ To this sequence of 3 exchange 
gates we now add a fourth exchange gate, namely $\exp{( it_4E_{12} )}$
in order 
to allow for an arbitrary phase to be obtained when the state is
located at
the north pole of the Bloch sphere. This extra gate corresponds to the first $\sigma_z$
rotation when a similar decomposition is considered for the Euler angle
construction for rotations in $SU(2)$ (Eq.~\ref{eqn:euler}). We now
use the quaternion approach to find explicit exchange sequences for
several elementary gates.

{\em i) Similarity transformation for exact CNOT, S.} In
Section~\ref{subsec:exactcnot}) it was shown that $U_{cnot}^{exchange}$
was diagonalized as $D=S^{\dagger}U_{cnot}^{exchange}S$. Here
$S= I \otimes \frac{\sqrt{3}}{2}I - \frac{i}{2}\sigma_y$ which
corresponds to a rotation of $60^o$ about the y axis on the second
qubit. The action of $S$ on the second qubit can be
written as $cos(\frac{\pi/3}{2})I-i \,
sin(\frac{\pi/3}{2})\sigma_y$. Referring to Eq.~\ref{eqn:quaternionhyper}, it
is evident that in the quaternion representation of rotations in
$SO(3)$ this corresponds to 
\begin{equation}
q = \{ \frac{\sqrt{3}}{2},\,0,\, \frac{1}{2},\, 0 \}
\label{eqn:qs}
\end{equation}
For sequential application of two exchange gates,
$\exp{(it_2E_{12})}\exp{(it_1E_{23})}$ or
$\exp{(it_2E_{23})} \exp{(it_1E_{12})}$, consideration of the
trigonometric equations that define the different 
components of the resulting quaternions shows that with
just two exchange gates it is impossible to satisfy the requirement that
both the $x$ component and $z$ component of the resulting quaternion be simultaneously zero
(Eq.~\ref{eqn:qs}). However, with a
sequence of three exchange gates ($\exp{( it_3E_{12})} \exp{(
it_2E_{23})} \exp{( it_1E_{12})}$), which in the
quaternion representation corresponds to 
\begin{eqnarray}
q_S(t_1,t_2,t_3) & = & \{ cos(t_2)cos(t_1+t_3) \ 
+ \frac{1}{2}sin(t_2)sin(t_1+t_3), \nonumber \\ 
& & \frac{\sqrt{3}}{2}sin(t_2)cos(t_1-t_3), \nonumber \\
& &  \frac{\sqrt{3}}{2}sin(t_2)sin(t_1-t_3), \nonumber \\
& & - cos(t_2)sin(t_1+t_3) \ 
+ \frac{1}{2}sin(t_2)cos(t_1+t_3) \},
\label{eqn:e12e23e12}
\end{eqnarray}
we find that a solution for $S$ can be obtained. This is accomplished as
follows. We need the $x$ component of $q_S$ to be zero
and the $y$ component non-zero (Eq.~\ref{eqn:qs}). Thus we set $t_1-t_3=\pi/2$
and solve for $t_2$ from 
\begin{equation}
\frac{\sqrt{3}}{2}sin(t_2)=\frac{1}{2}. 
\label{eqn:qS1}
\end{equation}
We then use the
first (scalar) component of $q_S$ to solve for $t_1+t_3$ from
\begin{equation}
\sqrt{\frac{2}{3}}cos(t_1+t_3) + \ 
\frac{1}{2\sqrt{3}}sin(t_1+t_3)  = 
\frac{\sqrt{3}}{2}, \nonumber \\
\end{equation}
which can be rewritten as
\begin{equation}
\frac{2}{3}\left( 1-\epsilon^2 \right)  = \left( \
\frac{\sqrt{3}}{2}-\frac{\epsilon}{2\sqrt{3}} \right)^2,
\label{eqn:qS2}
\end{equation}
where $\epsilon=sin(t_1+t_3)$. Solutions of Eqs.~\ref{eqn:qS1}
and~\ref{eqn:qS2} yields $t_1=asin(1/3)/2+\pi/4$, $t_2=asin(1/\sqrt{3})$, and
$t_3=asin(1/3)/2-\pi/4$. 

After obtaining the three-exchange representation of $S$, 
we still have to find the
single qubit gates that transforms the resulting matrix $D$ into the
C-PHASE gate. Since C-PHASE, like $D$, is diagonal in the
computational basis, a combination of $\sigma_z$ rotations and a
global phase is sufficient to transform one into the other. To find
the relevant rotation angles, $\phi$ and $\theta$ (Fig.~\ref{fig:fig5a}), and the global
phase factor, $\Omega$, we set up the following system of equations
\begin{eqnarray}
\Omega+\theta+\phi + \Delta_{00} & = & 0 \nonumber \\
\Omega+\theta-\phi + \Delta_{01} & = & 0 \nonumber \\
\Omega-\theta+\phi + \Delta_{10} & = & 0 \nonumber \\
\Omega-\theta-\phi + \Delta_{11} & = & \pi.
\end{eqnarray}
Here the $\Delta_{ij}$'s are the arguments of the diagonal matrix elements of $D$
when these are written as phases, {\em i.e.}
$\exp(i\Delta_{ij})=D_{ij}$, and the other terms on the left
are obtained from the diagonal elements of the matrix
$\exp{(i\Omega)}exp{( i (\theta \, \sigma_z \otimes I + \phi \, I
  \otimes \sigma_z) )}$. Solving for these variables, one finds that $\phi=0.612497$ and
$\theta=-0.547580$. To recast these in terms of exchange gates it is enough to realize that
implementing $E_{12}$ is equivalent to a $\sigma_z$ rotation (see
Eq.~\ref{eqn:exchangegates}). Thus, the exchange gate times
corresponding to $\theta$ and $\phi$ are $t_{\theta}=\theta$ and
$t_{\phi}=\phi$, respectively. For the $\pi/8$ gate the same argument
trivially yields a single exchange gate time of $t_{\pi/8}=\pi/8$. 

{\em ii) Hadamard gate.} Using the same sequence of three exchange
gates, $\exp{( it_3E_{12})} \exp{( it_2E_{23})} \exp{( it_1E_{12})}$,
and realizing that for the Hadamard gate it is the $y$
component of the resulting quaternion that must be zero, we can find
the solution for the Hadamard gate in an identical fashion. This
results in a quaternion representation:
\begin{equation}
q_H \left( t_1, \, t_2, \, t_3 \right) = q_H \left( -atan(\sqrt{2})/2, \, asin(\sqrt{2/3}), \,
-atan(\sqrt{2})/2 \right).
\end{equation}

{\em iii) NOT gate.} The NOT gate has a quaternion representation 
\begin{equation}
q_N = \{ 0,\,1,\, 0,\, 0 \}
\label{eqn:qn}
\end{equation} 
which corresponds to a full rotation from the south pole to north, or
vice versa, when interpreted on the Bloch sphere. For this gate the
three exchange gate sequence considered above is
insufficient to generate the gate, since this sequence can
never result in an $x$ component greater than $\sqrt{3}/2$ (see Eq.~\ref{eqn:e12e23e12}),
whereas the NOT gate requires an $x$ component of $1$. We can find a
solution for the NOT gate using the modified three exchange gate sequence $\exp{( it_3E_{23} )}
\exp{( it_2E_{12} )} \exp{( it_1E_{23} )}$. This sequence leads to
the following expression for the corresponding quaternion: 
\begin{eqnarray}
q_N(t_1,t_2,t_3) & = & \{ cos(t_2)cos(t_1+t_3) \ 
+ \frac{1}{2}sin(t_2)sin(t_1+t_3), \nonumber \\ 
& & \frac{\sqrt{3}}{2}cos(t_2)sin(t_1+t_3) \ 
+\frac{\sqrt{3}}{2}sin(t_1)sin(t_2)sin(t_3), \nonumber \\
& & - \frac{\sqrt{3}}{2}sin(t_2)sin(t_1-t_3), \nonumber \\
& & - \frac{1}{2}sin(t_2)cos(t_1-t_3) \ 
+\frac{1}{2}cos(t_2)sin(t_1+t_3) \nonumber \\
& & -\frac{1}{2}cos(t_1)sin(t_2)cos(t_3)  \},
\end{eqnarray}
We can then solve for the times $t_s$, $s=1\, -\, 3$, by first recognizing
that since the $y$ component must equal $0$ and the $x$ component
must equal $1$ then, $t_1-t_3=0$. Making use of this and the expressions for the $x$ and
$z$ components, we can then solve for $t_2$ from
$t_2=asin(1/\sqrt{3})$. Substituting this back into the expression for
the scalar component, which also must be equal to zero, we then
obtain the corresponding values for $t_1$ and $t_3$ as $t_1=t_3=atan(\sqrt{2})$.

\end{appendix}


\begin{thebibliography}{40}
\expandafter\ifx\csname natexlab\endcsname\relax\def\natexlab#1{#1}\fi
\expandafter\ifx\csname bibnamefont\endcsname\relax
  \def\bibnamefont#1{#1}\fi
\expandafter\ifx\csname bibfnamefont\endcsname\relax
  \def\bibfnamefont#1{#1}\fi
\expandafter\ifx\csname citenamefont\endcsname\relax
  \def\citenamefont#1{#1}\fi
\expandafter\ifx\csname url\endcsname\relax
  \def\url#1{\texttt{#1}}\fi
\expandafter\ifx\csname urlprefix\endcsname\relax\def\urlprefix{URL }\fi
\providecommand{\bibinfo}[2]{#2}
\providecommand{\eprint}[2][]{\url{#2}}

\bibitem[{\citenamefont{DiVincenzo et~al.}(2000)\citenamefont{DiVincenzo,
  Bacon, Kempe, Burkard, and Whaley}}]{divincenzo00}
\bibinfo{author}{\bibfnamefont{D.~P.} \bibnamefont{DiVincenzo}},
  \bibinfo{author}{\bibfnamefont{D.}~\bibnamefont{Bacon}},
  \bibinfo{author}{\bibfnamefont{J.}~\bibnamefont{Kempe}},
  \bibinfo{author}{\bibfnamefont{G.}~\bibnamefont{Burkard}}, \bibnamefont{and}
  \bibinfo{author}{\bibfnamefont{K.~B.} \bibnamefont{Whaley}},
  \bibinfo{journal}{Nature} \textbf{\bibinfo{volume}{408}},
  \bibinfo{pages}{339} (\bibinfo{year}{2000}).

\bibitem[{\citenamefont{Kempe et~al.}(2001)\citenamefont{Kempe, Bacon, Lidar,
  and Whaley}}]{kempe01}
\bibinfo{author}{\bibfnamefont{J.}~\bibnamefont{Kempe}},
  \bibinfo{author}{\bibfnamefont{D.}~\bibnamefont{Bacon}},
  \bibinfo{author}{\bibfnamefont{D.~A.} \bibnamefont{Lidar}}, \bibnamefont{and}
  \bibinfo{author}{\bibfnamefont{K.~B.} \bibnamefont{Whaley}},
  \bibinfo{journal}{Phys.\ Rev.\ A} \textbf{\bibinfo{volume}{63}},
  \bibinfo{pages}{042037} (\bibinfo{year}{2001}).

\bibitem[{\citenamefont{Bacon et~al.}(2000)\citenamefont{Bacon, Kempe, Lidar,
  and Whaley}}]{bacon00}
\bibinfo{author}{\bibfnamefont{D.}~\bibnamefont{Bacon}},
  \bibinfo{author}{\bibfnamefont{J.}~\bibnamefont{Kempe}},
  \bibinfo{author}{\bibfnamefont{D.}~\bibnamefont{Lidar}}, \bibnamefont{and}
  \bibinfo{author}{\bibfnamefont{K.}~\bibnamefont{Whaley}},
  \bibinfo{journal}{Phys.\ Rev.\ Lett.} \textbf{\bibinfo{volume}{85}},
  \bibinfo{pages}{1758} (\bibinfo{year}{2000}).

\bibitem[{\citenamefont{Landauer}(1996)}]{landauer96}
\bibinfo{author}{\bibfnamefont{R.}~\bibnamefont{Landauer}},
  \bibinfo{journal}{Science} \textbf{\bibinfo{volume}{272}},
  \bibinfo{pages}{1914} (\bibinfo{year}{1996}).

\bibitem[{\citenamefont{Barenco et~al.}(1995)\citenamefont{Barenco, Deutsch,
  and Ekert}}]{barenco95}
\bibinfo{author}{\bibfnamefont{A.}~\bibnamefont{Barenco}},
  \bibinfo{author}{\bibfnamefont{D.}~\bibnamefont{Deutsch}}, \bibnamefont{and}
  \bibinfo{author}{\bibfnamefont{A.}~\bibnamefont{Ekert}},
  \bibinfo{journal}{Phys.\ Rev.\ Lett.} \textbf{\bibinfo{volume}{74}},
  \bibinfo{pages}{4083} (\bibinfo{year}{1995}).

\bibitem[{\citenamefont{Loss and DiVincenzo}(1998)}]{loss98}
\bibinfo{author}{\bibfnamefont{D.}~\bibnamefont{Loss}} \bibnamefont{and}
  \bibinfo{author}{\bibfnamefont{D.~P.} \bibnamefont{DiVincenzo}},
  \bibinfo{journal}{Phys.\ Rev.\ A} \textbf{\bibinfo{volume}{57}},
  \bibinfo{pages}{120} (\bibinfo{year}{1998}).

\bibitem[{\citenamefont{Loss et~al.}(1998)\citenamefont{Loss, DiVincenzo, and
  Loss}}]{divincenzo98}
\bibinfo{author}{\bibfnamefont{D.}~\bibnamefont{Loss}},
  \bibinfo{author}{\bibfnamefont{D.~P.} \bibnamefont{DiVincenzo}},
  \bibnamefont{and} \bibinfo{author}{\bibfnamefont{D.}~\bibnamefont{Loss}},
  \bibinfo{journal}{Superlattices and Microstructures}
  \textbf{\bibinfo{volume}{23}}, \bibinfo{pages}{419} (\bibinfo{year}{1998}).

\bibitem[{\citenamefont{Burkard et~al.}(1999)\citenamefont{Burkard, Loss, and
  DiVincenzo}}]{burkard99}
\bibinfo{author}{\bibfnamefont{G.}~\bibnamefont{Burkard}},
  \bibinfo{author}{\bibfnamefont{D.}~\bibnamefont{Loss}}, \bibnamefont{and}
  \bibinfo{author}{\bibfnamefont{D.~P.} \bibnamefont{DiVincenzo}},
  \bibinfo{journal}{Phys.\ Rev.\ B} \textbf{\bibinfo{volume}{59}},
  \bibinfo{pages}{2070} (\bibinfo{year}{1999}).

\bibitem[{\citenamefont{Kikkawa et~al.}(1997)\citenamefont{Kikkawa, Smorchkova,
  Samarth, and Awschalom}}]{kikkawa97}
\bibinfo{author}{\bibfnamefont{J.~M.} \bibnamefont{Kikkawa}},
  \bibinfo{author}{\bibfnamefont{I.~P.} \bibnamefont{Smorchkova}},
  \bibinfo{author}{\bibfnamefont{N.}~\bibnamefont{Samarth}}, \bibnamefont{and}
  \bibinfo{author}{\bibfnamefont{D.~D.} \bibnamefont{Awschalom}},
  \bibinfo{journal}{Science} \textbf{\bibinfo{volume}{277}},
  \bibinfo{pages}{1284} (\bibinfo{year}{1997}).

\bibitem[{\citenamefont{Fujisawa et~al.}(2001)\citenamefont{Fujisawa, Tokura,
  and Hirayama}}]{fujisawa01}
\bibinfo{author}{\bibfnamefont{T.}~\bibnamefont{Fujisawa}},
  \bibinfo{author}{\bibfnamefont{Y.}~\bibnamefont{Tokura}}, \bibnamefont{and}
  \bibinfo{author}{\bibfnamefont{Y.}~\bibnamefont{Hirayama}},
  \bibinfo{journal}{Phys.\ Rev.\ B} \textbf{\bibinfo{volume}{63}},
  \bibinfo{pages}{081304} (\bibinfo{year}{2001}).

\bibitem[{\citenamefont{Livermore et~al.}(1996)\citenamefont{Livermore, Crouch,
  Westervelt, Campman, and Gossard}}]{livermore96}
\bibinfo{author}{\bibfnamefont{C.}~\bibnamefont{Livermore}},
  \bibinfo{author}{\bibfnamefont{C.~H.} \bibnamefont{Crouch}},
  \bibinfo{author}{\bibfnamefont{R.~M.} \bibnamefont{Westervelt}},
  \bibinfo{author}{\bibfnamefont{K.~L.} \bibnamefont{Campman}},
  \bibnamefont{and} \bibinfo{author}{\bibfnamefont{A.~C.}
  \bibnamefont{Gossard}}, \bibinfo{journal}{Science}
  \textbf{\bibinfo{volume}{274}}, \bibinfo{pages}{1332} (\bibinfo{year}{1996}).

\bibitem[{\citenamefont{Tarucha et~al.}(1996)\citenamefont{Tarucha, Austing,
  Honda, van~der Hage, and Kouwenhoven}}]{tarucha96}
\bibinfo{author}{\bibfnamefont{S.}~\bibnamefont{Tarucha}},
  \bibinfo{author}{\bibfnamefont{D.~G.} \bibnamefont{Austing}},
  \bibinfo{author}{\bibfnamefont{T.}~\bibnamefont{Honda}},
  \bibinfo{author}{\bibfnamefont{R.~J.} \bibnamefont{van~der Hage}},
  \bibnamefont{and} \bibinfo{author}{\bibfnamefont{L.~P.}
  \bibnamefont{Kouwenhoven}}, \bibinfo{journal}{Phys.\ Rev.\ Lett.}
  \textbf{\bibinfo{volume}{77}}, \bibinfo{pages}{3613} (\bibinfo{year}{1996}).

\bibitem[{\citenamefont{Gupta et~al.}(1999)\citenamefont{Gupta, Awschalom,
  Peng, and Alivisatos}}]{gupta99}
\bibinfo{author}{\bibfnamefont{J.~A.} \bibnamefont{Gupta}},
  \bibinfo{author}{\bibfnamefont{D.~D.} \bibnamefont{Awschalom}},
  \bibinfo{author}{\bibfnamefont{X.}~\bibnamefont{Peng}}, \bibnamefont{and}
  \bibinfo{author}{\bibfnamefont{A.~P.} \bibnamefont{Alivisatos}},
  \bibinfo{journal}{Phys.\ Rev.\ B} \textbf{\bibinfo{volume}{59}},
  \bibinfo{pages}{10421} (\bibinfo{year}{1999}).

\bibitem[{\citenamefont{Wiseman et~al.}(2001)\citenamefont{Wiseman, Utami, Sun,
  Milburn, Kane, Dzurak, and Clark}}]{wiseman01}
\bibinfo{author}{\bibfnamefont{H.}~\bibnamefont{Wiseman}},
  \bibinfo{author}{\bibfnamefont{D.~W.} \bibnamefont{Utami}},
  \bibinfo{author}{\bibfnamefont{H.~B.} \bibnamefont{Sun}},
  \bibinfo{author}{\bibfnamefont{G.}~\bibnamefont{Milburn}},
  \bibinfo{author}{\bibfnamefont{B.~E.} \bibnamefont{Kane}},
  \bibinfo{author}{\bibfnamefont{A.}~\bibnamefont{Dzurak}}, \bibnamefont{and}
  \bibinfo{author}{\bibfnamefont{R.~G.} \bibnamefont{Clark}},
  \bibinfo{journal}{Phys.\ Rev.\ B} \textbf{\bibinfo{volume}{63}},
  \bibinfo{pages}{235308} (\bibinfo{year}{2001}).

\bibitem[{\citenamefont{Engel and Loss}(2001)}]{engel01}
\bibinfo{author}{\bibfnamefont{H.-A.} \bibnamefont{Engel}} \bibnamefont{and}
  \bibinfo{author}{\bibfnamefont{D.}~\bibnamefont{Loss}},
  \bibinfo{journal}{Phys.\ Rev.\ Lett.} \textbf{\bibinfo{volume}{86}},
  \bibinfo{pages}{4648} (\bibinfo{year}{2001}).

\bibitem[{\citenamefont{DiVincenzo}(1999)}]{divincenzo99b}
\bibinfo{author}{\bibfnamefont{D.~P.} \bibnamefont{DiVincenzo}},
  \bibinfo{journal}{Jour. Appl. Phys.} \textbf{\bibinfo{volume}{85}},
  \bibinfo{pages}{4785} (\bibinfo{year}{1999}).

\bibitem[{\citenamefont{Vala and Whaley}(2002)}]{vala02}
\bibinfo{author}{\bibfnamefont{J.}~\bibnamefont{Vala}} \bibnamefont{and}
  \bibinfo{author}{\bibfnamefont{K.~B.} \bibnamefont{Whaley}},
  \bibinfo{journal}{Phys.\ Rev.\ A} \textbf{\bibinfo{volume}{66}},
  \bibinfo{pages}{022304} (\bibinfo{year}{2002}).

\bibitem[{\citenamefont{Bonesteel et~al.}(2001)\citenamefont{Bonesteel,
  Stepanenko, and DiVincenzo}}]{bonesteel01}
\bibinfo{author}{\bibfnamefont{N.}~\bibnamefont{Bonesteel}},
  \bibinfo{author}{\bibfnamefont{D.}~\bibnamefont{Stepanenko}},
  \bibnamefont{and}
  \bibinfo{author}{\bibfnamefont{D.}~\bibnamefont{DiVincenzo}},
  \bibinfo{journal}{Phys.\ Rev.\ Lett.} \textbf{\bibinfo{volume}{87}},
  \bibinfo{pages}{207901} (\bibinfo{year}{2001}).

\bibitem[{\citenamefont{Makhlin}(2002)}]{makhlin02}
\bibinfo{author}{\bibfnamefont{Y.}~\bibnamefont{Makhlin}},
  \bibinfo{journal}{Quant. Info. Proc.} \textbf{\bibinfo{volume}{1}},
  \bibinfo{pages}{243} (\bibinfo{year}{2002}).

\bibitem[{\citenamefont{Nielsen and Chuang}(2000)}]{nielsen00}
\bibinfo{author}{\bibfnamefont{M.}~\bibnamefont{Nielsen}} \bibnamefont{and}
  \bibinfo{author}{\bibfnamefont{I.}~\bibnamefont{Chuang}},
  \emph{\bibinfo{title}{Quantum Computation and Quantum Information}}
  (\bibinfo{publisher}{Cambridge University Press},
  \bibinfo{address}{Cambridge, UK}, \bibinfo{year}{2000}).

\bibitem[{\citenamefont{Nelder and Mead}(1965)}]{nelder65}
\bibinfo{author}{\bibfnamefont{J.}~\bibnamefont{Nelder}} \bibnamefont{and}
  \bibinfo{author}{\bibfnamefont{R.}~\bibnamefont{Mead}},
  \bibinfo{journal}{Computer Journal} \textbf{\bibinfo{volume}{7}},
  \bibinfo{pages}{308} (\bibinfo{year}{1965}).

\bibitem[{\citenamefont{Lagarias et~al.}(1995)\citenamefont{Lagarias, Reeds,
  Wright, and Wright}}]{lagarias95}
\bibinfo{author}{\bibfnamefont{J.~C.} \bibnamefont{Lagarias}},
  \bibinfo{author}{\bibfnamefont{J.~A.} \bibnamefont{Reeds}},
  \bibinfo{author}{\bibfnamefont{M.~H.} \bibnamefont{Wright}},
  \bibnamefont{and} \bibinfo{author}{\bibfnamefont{P.~E.}
  \bibnamefont{Wright}}, \bibinfo{journal}{SIAM J. Optim.}
  \textbf{\bibinfo{volume}{9}}, \bibinfo{pages}{113} (\bibinfo{year}{1995}).

\bibitem[{\citenamefont{Sakurai}(1994)}]{sakurai94}
\bibinfo{author}{\bibfnamefont{J.}~\bibnamefont{Sakurai}},
  \emph{\bibinfo{title}{Modern Quantum Mechanics}}
  (\bibinfo{publisher}{Addison-Wesley}, \bibinfo{address}{US},
  \bibinfo{year}{1994}).

\bibitem[{\citenamefont{Dalibard et~al.}(1992)\citenamefont{Dalibard, Castin,
  and M{\o}lmer}}]{dalibard92}
\bibinfo{author}{\bibfnamefont{J.}~\bibnamefont{Dalibard}},
  \bibinfo{author}{\bibfnamefont{Y.}~\bibnamefont{Castin}}, \bibnamefont{and}
  \bibinfo{author}{\bibfnamefont{K.}~\bibnamefont{M{\o}lmer}},
  \bibinfo{journal}{Phys.\ Rev.\ Lett.} \textbf{\bibinfo{volume}{68}},
  \bibinfo{pages}{580} (\bibinfo{year}{1992}).

\bibitem[{\citenamefont{M{\o}lmer et~al.}(1993)\citenamefont{M{\o}lmer, Castin,
  and Dalibard}}]{molmer93}
\bibinfo{author}{\bibfnamefont{K.}~\bibnamefont{M{\o}lmer}},
  \bibinfo{author}{\bibfnamefont{Y.}~\bibnamefont{Castin}}, \bibnamefont{and}
  \bibinfo{author}{\bibfnamefont{J.}~\bibnamefont{Dalibard}},
  \bibinfo{journal}{J.\ Opt.\ Soc.\ Am.\ B} \textbf{\bibinfo{volume}{10}},
  \bibinfo{pages}{524} (\bibinfo{year}{1993}).

\bibitem[{\citenamefont{Hegerfeldt and Wilser}(1991)}]{hegerfeldt91}
\bibinfo{author}{\bibfnamefont{G.~C.} \bibnamefont{Hegerfeldt}}
  \bibnamefont{and} \bibinfo{author}{\bibfnamefont{T.~S.}
  \bibnamefont{Wilser}}, in \emph{\bibinfo{booktitle}{Proceedings of the Second
  International Wigner Symposium}}, edited by
  \bibinfo{editor}{\bibfnamefont{H.}~\bibnamefont{Doebner}},
  \bibinfo{editor}{\bibfnamefont{W.}~\bibnamefont{Scherer}}, \bibnamefont{and}
  \bibinfo{editor}{\bibfnamefont{J.}~\bibnamefont{F.~Schroeck}}
  (\bibinfo{publisher}{World Scientific}, \bibinfo{address}{Singapore},
  \bibinfo{year}{1991}), p. \bibinfo{pages}{104}.

\bibitem[{\citenamefont{Carmichael}(1993)}]{carmichael93}
\bibinfo{author}{\bibfnamefont{H.}~\bibnamefont{Carmichael}},
  \emph{\bibinfo{title}{An open systems approach to quantum optics}}, Lecture
  notes in physics (\bibinfo{publisher}{Springer}, \bibinfo{address}{Berlin},
  \bibinfo{year}{1993}).

\bibitem[{\citenamefont{Awschalom and Kikkawa}(1999)}]{awschalom99}
\bibinfo{author}{\bibfnamefont{D.~D.} \bibnamefont{Awschalom}}
  \bibnamefont{and} \bibinfo{author}{\bibfnamefont{J.~M.}
  \bibnamefont{Kikkawa}}, \bibinfo{journal}{Phys. Today}
  \textbf{\bibinfo{volume}{52}}, \bibinfo{pages}{33} (\bibinfo{year}{1999}).

\bibitem[{\citenamefont{Zhang and Whaley}(1991)}]{zhang91a}
\bibinfo{author}{\bibfnamefont{Q.}~\bibnamefont{Zhang}} \bibnamefont{and}
  \bibinfo{author}{\bibfnamefont{K.~B.} \bibnamefont{Whaley}},
  \bibinfo{journal}{Phys.\ Rev.\ B} \textbf{\bibinfo{volume}{43}},
  \bibinfo{pages}{11062} (\bibinfo{year}{1991}).

\bibitem[{\citenamefont{Zhang and Whaley}(1992)}]{zhang92}
\bibinfo{author}{\bibfnamefont{Q.}~\bibnamefont{Zhang}} \bibnamefont{and}
  \bibinfo{author}{\bibfnamefont{K.~B.} \bibnamefont{Whaley}},
  \bibinfo{journal}{Jour. Chem. Phys.} \textbf{\bibinfo{volume}{96}},
  \bibinfo{pages}{5318} (\bibinfo{year}{1992}).

\bibitem[{\citenamefont{Hahn}(1950)}]{hahn50}
\bibinfo{author}{\bibfnamefont{E.}~\bibnamefont{Hahn}}, \bibinfo{journal}{Phys.
  Rev.} \textbf{\bibinfo{volume}{80}}, \bibinfo{pages}{580}
  (\bibinfo{year}{1950}).

\bibitem[{\citenamefont{Bacon et~al.}(1999)\citenamefont{Bacon, Lidar, and
  Whaley}}]{bacon99}
\bibinfo{author}{\bibfnamefont{D.}~\bibnamefont{Bacon}},
  \bibinfo{author}{\bibfnamefont{D.}~\bibnamefont{Lidar}}, \bibnamefont{and}
  \bibinfo{author}{\bibfnamefont{K.}~\bibnamefont{Whaley}},
  \bibinfo{journal}{Phys.\ Rev.\ A} \textbf{\bibinfo{volume}{60}},
  \bibinfo{pages}{1944} (\bibinfo{year}{1999}).

\bibitem[{\citenamefont{Sanders et~al.}(1999)\citenamefont{Sanders, Kim, and
  Holton}}]{sanders99}
\bibinfo{author}{\bibfnamefont{G.}~\bibnamefont{Sanders}},
  \bibinfo{author}{\bibfnamefont{K.}~\bibnamefont{Kim}}, \bibnamefont{and}
  \bibinfo{author}{\bibfnamefont{W.}~\bibnamefont{Holton}},
  \bibinfo{journal}{Phys.\ Rev.\ A} \textbf{\bibinfo{volume}{59}},
  \bibinfo{pages}{1098} (\bibinfo{year}{1999}).

\bibitem[{\citenamefont{Niskanen et~al.}(2003)\citenamefont{Niskanen,
  Vartianen, and Salomaa}}]{niskanen03}
\bibinfo{author}{\bibfnamefont{A.~O.} \bibnamefont{Niskanen}},
  \bibinfo{author}{\bibfnamefont{J.~J.} \bibnamefont{Vartianen}},
  \bibnamefont{and} \bibinfo{author}{\bibfnamefont{M.~M.}
  \bibnamefont{Salomaa}}, \bibinfo{journal}{Phys.\ Rev.\ Lett.}
  \textbf{\bibinfo{volume}{90}}, \bibinfo{pages}{197901}
  (\bibinfo{year}{2003}).

\bibitem[{\citenamefont{Chen et~al.}(2001)\citenamefont{Chen, Piermarocchi, and
  Sham}}]{chen01a}
\bibinfo{author}{\bibfnamefont{P.}~\bibnamefont{Chen}},
  \bibinfo{author}{\bibfnamefont{C.}~\bibnamefont{Piermarocchi}},
  \bibnamefont{and} \bibinfo{author}{\bibfnamefont{L.~J.} \bibnamefont{Sham}},
  \bibinfo{journal}{Physica E} \textbf{\bibinfo{volume}{10}},
  \bibinfo{pages}{7} (\bibinfo{year}{2001}).

\bibitem[{\citenamefont{Hamilton}(1967)}]{hamilton67}
\bibinfo{author}{\bibfnamefont{W.}~\bibnamefont{Hamilton}},
  \emph{\bibinfo{title}{The Mathematical Papers of Sir William Rowan
  Hamilton.}} (\bibinfo{publisher}{Cambridge University Press},
  \bibinfo{address}{Cambridge, UK}, \bibinfo{year}{1967}).

\bibitem[{\citenamefont{Kuipers}(1998)}]{kuipers98}
\bibinfo{author}{\bibfnamefont{J.}~\bibnamefont{Kuipers}},
  \emph{\bibinfo{title}{Quaternions and Rotation Sequences: A Primer with
  Applications to Orbits, Aerospace, and Virtual Reality}}
  (\bibinfo{publisher}{Princeton University Press},
  \bibinfo{address}{Princeton, NJ}, \bibinfo{year}{1998}).

\bibitem[{\citenamefont{Zare}(1988)}]{zare88}
\bibinfo{author}{\bibfnamefont{R.~N.} \bibnamefont{Zare}},
  \emph{\bibinfo{title}{Angular Momentum}} (\bibinfo{publisher}{John Wiley \&
  Sons, Inc.}, \bibinfo{address}{US}, \bibinfo{year}{1988}).

\bibitem[{\citenamefont{D'Alessandro}(2001)}]{dalessandro01}
\bibinfo{author}{\bibfnamefont{D.}~\bibnamefont{D'Alessandro}},
  \bibinfo{journal}{quant-ph/0110120}  (\bibinfo{year}{2001}).

\bibitem[{\citenamefont{Zhang}(2003)}]{junzhang03}
\bibinfo{author}{\bibfnamefont{J.}~\bibnamefont{Zhang}}, Ph.D. thesis,
  \bibinfo{school}{University of California, Berkeley} (\bibinfo{year}{2003}).

\end{thebibliography}

\newpage

\begingroup

\begin{table}[t]
\caption{Exchange gate times for the full sequence for CNOT in the
computational basis, Eq.~\ref{eq:CNOT_exchange}, determined by
numerical optimization. Exchanges 1-8 correspond to 
  $U_1 \otimes U_2$, exchanges 9-27 
  to $U_{cnot}^{exchange}$, and exchanges 28-35 to
  $V_1 \otimes V_2$ (see Fig.~\ref{fig:fig4}). 
Starting from the gate times for exchanges 9-27 
  given in Ref.~\onlinecite{divincenzo00}, a Nelder-Mead optimization
  routine was used to find the additional gate times corresponding to gates 1-8
  and 28-35. The optimization criteria was minimization of the 
  the cost function of
  Eq.~\ref{eqn:costfunction} over the 16-dimensional parameter space
of exchanges 1-8 and 28-35.
 The sequence of 35 gates and associated
  gate times seen here represent the first 
  successful run where the cost function dropped below the required
  tolerance of $10^{-4}$.
 All gate times are given as positive numbers modulo $\pi$, in units
of $2\hbar/J_0$. }   
\label{tab:tab1}
\begin{tabular}{|c|c|c|c|c|c|c|c|}\hline 
Exchange & Qubit  & Qubit &          & Exchange & Qubit & Qubit & \\
Number   &    1   &   2   & Time     & Number   &   1   &   2   & Time \\
\hline \hline
1        &    1   &   2   & 2.462204 & 19       &    2   &   3   & 1.302882 \\ 
2        &    2   &   3   & 0.977712 & 20       &    3   &   4   & 0.463868 \\
3        &    1   &   2   & 2.209031 & 21       &    2   &   3   & 2.554511 \\
4        &    2   &   3   & 0.977711 & 22       &    4   &   5   & 0.871873 \\ 
5        &    4   &   5   & 0.690514 & 23       &    1   &   2   & 1.249644 \\
6        &    5   &   6   & 2.837899 & 24       &    5   &   6   & 2.107472 \\
7        &    4   &   5   & 2.298306 & 25       &    2   &   3   & 2.554511 \\
8        &    5   &   6   & 1.411241 & 26       &    4   &   5   & 0.871873 \\

9        &    3   &   4   & 1.290877 & 27       &    3   &   4   & 1.290877 \\
10       &    2   &   3   & 0.650655 & 28       &    1   &   2   & 0.727495 \\
11       &    4   &   5   & 0.871873 & 29       &    2   &   3   & 1.761338 \\
12       &    1   &   2   & 1.934484 & 30       &    1   &   2   & 0.368173 \\
13       &    5   &   6   & 2.107472 & 31       &    2   &   3   & 1.761338 \\   
14       &    2   &   3   & 0.650656 & 32       &    4   &   5   & 2.820908 \\
15       &    4   &   5   & 0.871873 & 33       &    5   &   6   & 3.709248 \\
16       &    3   &   4   & 2.012206 & 34       &    4   &   5   & 0.090528 \\
17       &    2   &   3   & 1.302882 & 35       &    5   &   6   & 1.622010 \\
18       &    1   &   2   & 2.639495 &          &        &       &          \\
\hline
\end{tabular}
\end{table}

\newpage

\begin{table}[t]
\caption{Exchange gates and times required to
transform the exchange CNOT gate into the exact CNOT gate in the
computational basis, Eq.~(\ref{eq:CNOT_exchange}), obtained with 
analytic solution
of the local transformations $U_i, V_i, i=1,2$. 
Exchange gates 1-6 correspond to $U_1 \otimes U_2$, exchange gates 7-25 to
$U_{cnot}^{exchange}$, and exchange gates 26-30 to $V_1 \otimes V_2$ (see
Fig.~\ref{fig:fig6}). 
Exchanges 7-25 are taken from Ref.~\onlinecite{divincenzo00}. The
remaining gates (1-6 and 26-30) were arrived at analytically by
decomposing each local unitary into 
a sequence of simpler rotations as described in the text, and then using the 
quaternion
representation to find the corresponding rotations in $SO(3)$. See
Appendix~\ref{app:quaternions} for full details. All gate times are given as positive numbers modulo
$\pi$, in units of $2\hbar/J_0$. }       
\label{tab:tab2}
\begin{tabular}{|c|c|c|c|c|c|c|c|}\hline 
Exchange & Qubit  & Qubit &           & Exchange & Qubit & Qubit & \\
Number   &    1   &   2   & Time      & Number   &    1  &   2   & Time \\
\hline \hline
1        &    4   &   5   & 2.663935  & 16       &    1   &   2   & 2.639495 \\ 
2        &    5   &   6   & 0.955317  & 17       &    2   &   3   & 1.302882 \\
3        &    1   &   2   & 0.612498  & 18       &    3   &   4   & 0.463868 \\

4        &    4   &   5   & 1.161038  & 19       &    2   &   3   & 2.554511 \\
5        &    5   &   6   & 2.526113  & 20       &    4   &   5   & 0.871873 \\
6        &    4   &   5   & 0.615480  & 21       &    1   &   2   & 1.249644 \\

7        &    3   &   4   & 1.290877  & 22       &    5   &   6   & 2.107472 \\
8        &    2   &   3   & 0.650655  & 23       &    2   &   3   & 2.554511 \\
9        &    4   &   5   & 0.871873  & 24       &    4   &   5   & 0.871873 \\
10       &    1   &   2   & 1.934484  & 25       &    3   &   4   & 1.290877 \\
11       &    5   &   6   & 2.107472  & 26       &    4   &   5   & 2.526113 \\
12       &    2   &   3   & 0.650656  & 27       &    5   &   6   & 0.615480 \\
13       &    4   &   5   & 0.871873  & 28       &    4   &   5   & 0.477659 \\
14       &    3   &   4   & 2.012206  & 29       &    5   &   6   & 0.955317 \\ 
15       &    2   &   3   & 1.302882  & 30       &    4   &   5   & 2.663935 \\
\hline
\end{tabular}
\end{table}
  
\newpage

\begin{table}[t]
\caption{One possible set of exchange gates and times that
implements the two-qubit unitary of a CNOT sandwiched between four
Hadamards (Fig.~\ref{fig:fig14}, equivalent to a CNOT with target and control
qubits reversed). 
Starting with the 19-gate exchange-only CNOT sequence for $U_{cnot}^{exchange}$
of Ref.~\onlinecite{divincenzo00}, the additional local gates were then
determined analytically using the 
quaternion formulation summarized in Appendix~\ref{app:quaternions}.
All gate times are given as positive numbers modulo $\pi$, in units of
$2\hbar/J_0$.} 
\label{tab:tab3}
\begin{tabular}{|c|c|c|c|c|c|c|c|}\hline 
Exchange & Qubit  & Qubit &           & Exchange & Qubit & Qubit & \\
Number   &    1   &   2   & Time      & Number   &    1  &   2   & Time \\
\hline \hline
1        &    4   &   5   & 1.638696 & 17       &    2   &   3   & 1.302882 \\
2        &    5   &   6   & 2.526113 & 18       &    3   &   4   & 0.463868 \\
3        &    4   &   5   & 0.615480 & 19       &    2   &   3   & 2.554511 \\
4        &    1   &   2   & 2.663935 & 20       &    4   &   5   & 0.871873 \\
5        &    2   &   3   & 0.955317 & 21       &    1   &   2   & 1.249644 \\
6        &    1   &   2   & 0.134839 & 22       &    5   &   6   & 2.107472 \\

7        &    3   &   4   & 1.290877 & 23       &    2   &   3   & 2.554511 \\
8        &    2   &   3   & 0.650655 & 24       &    4   &   5   & 0.871873 \\
9        &    4   &   5   & 0.871873 & 25       &    3   &   4   & 1.290877 \\
10       &    1   &   2   & 1.934484 & 26       &    4   &   5   & 2.526113 \\
11       &    5   &   6   & 2.107472 & 27       &    5   &   6   & 0.615480 \\
12       &    2   &   3   & 0.650656 & 28       &    4   &   5   & 0.955317 \\
13       &    4   &   5   & 0.871873 & 29       &    1   &   2   & 2.663935 \\
14       &    3   &   4   & 2.012206 & 30       &    2   &   3   & 0.955317 \\
15       &    2   &   3   & 1.302882 & 31       &    1   &   2   & 2.663935 \\
16       &    1   &   2   & 2.639495 &          &        &       & \\
\hline
\end{tabular}
\end{table}

\endgroup

\end{document}